\documentclass[10pt]{iopart}
\pdfoutput=1
\usepackage{graphicx}
\usepackage{setstack}
\usepackage{iopams}
 
\usepackage{color}
\usepackage{ulem}

\newcommand{\be}{\begin{equation}}
\newcommand{\ee}{\end{equation}}
\newcommand{\bea}{\begin{eqnarray}}
\newcommand{\eea}{\end{eqnarray}}
\newcommand{\im}{\mathrm{i}}

\begin{document}
\title{Excitonic effects in the optical properties of 2D materials: An equation of motion approach}

\author{A. J. Chaves$^1$, R. M. Ribeiro$^1$, T. Frederico$^2$, N. M. R. Peres$^1$}
\ead{andrej6@gmail.com}
\ead{ricardo@fisica.uminho.pt}
\ead{tobias@ita.br}
\ead{peres@fisica.uminho.pt}
\address{$^1$ Department of Physics and Center of Physics, and 
	QuantaLab, University of Minho, 4710-057, Braga, Portugal}
\address{$^2$ Department of Physics, Instituto Tecnol\'ogico de Aeron\'autica,
DCTA, 12228-900 S\~ao Jos\'e dos Campos, Brazil}

\date{\today}

\begin{abstract}
We present a unified description of the excitonic properties of four monolayer transition-metal dichalcogenides (TMDC's)
using an equation of motion method for deriving the Bethe-Salpeter equation in momentum space. 
Our method is able to cope with both continuous and tight-binding Hamiltonians, and is less computational demanding than the traditional first-principles approach.
We show that the role of the exchange energy is essential to 
obtain a good description of the binding energy of the excitons. 
The exchange energy at the $\Gamma-$point is also essential to obtain the correct position of the C-exciton peak.
Using our model we obtain a good agreement between the Rydberg series measured for WS$_2$. 
We discuss how the  absorption and the Rydberg series depend on the doping.
Choosing $r_0$ and the doping we obtain a good qualitative agreement between the experimental absorption and 
our calculations for MoS$_2$ and WS$_2$.
We also derive a semi-analytical version of Ellitot's formula for TMDC's.
\end{abstract}
\maketitle
\section{Introduction}

The study of excitons in bulk transition-metal dichalcogenides (TMDC's) is a research topic
in condensed matter physics that dates back to 1960's \cite{Frindt1965,wilson1969transition}. 
With the advent of two-dimensional materials \cite{novoselov2004electric}, this topic regained interest since 
it became possible to study single- and few-layers of TMDC's \cite{Potemski2016,mak2010atomically}. 
Together with its two-dimensional nature, this new class of materials also has a hexagonal lattice structure as does graphene.
On the other hand,
 while the low-energy electronic excitations are  in graphene described by a massless Dirac equation, in monolayer TMDC's the same excitations can be described by a massive (with a gap) Dirac equation. The absence of a gap in graphene prohibits the existence of bound-states of excitons (but not of excitonic resonances \cite{Peres2010excitonic}).
 On the contrary, we find in TMDC's absorption spectrum fingerprints of both excitonic bound states (including the presence of a Rydberg series)
 and of excitonic resonances, due to electron-hole scattering processes, with energies above the non-interacting gap.

As a consequence of optical experimental studies in few-layers TMDC's,
the study of a new type of excitons in these novel 2D materials 
became possible. This
has attracted a wealth of scientific research \cite{molina2016temperature,thilagam2014exciton,lundt2016room,Morozov2015,li2014measurement,Berkelbach2013,berkelbach2015,gao2016dynamical,molina2013effect,
qiu2013optical,berghauser2014analytical,Wu2015,zhou2015berry,srivastava2015signatures,MoS2exp1,steinhoff2014influence}. 
The signature of excitons appeared first in the optical
measurements of monolayer MoS$_2$ \cite{mak2010atomically}, where two peaks in the  absorbance, with energies $\sim1.9$ eV and $\sim2.1$ eV, were identified.
These two peaks
correspond approximately to the same results  found in several layers of MoS$_2$ \cite{Frindt1965}. 

The optical studies of other monolayers of TMDC's soon followed at the pace of their synthesis. The
 optical properties of MX$_2$, M=\{Mo,W\}, X=\{S,Se\}, in the range $1.5-3$ eV were experimentally studied by Li \textit{et al.} \cite{li2014measurement}, 
with reflectance and transmittance measurements followed by a 
Kramers-Kronig analyses, and by Morozov and Kuno \cite{Morozov2015}, with differential transmission and reflectance measurements.  
It should be noted that all these four materials have similar optical properties. Their optical absorbance spectra show signatures of the  spin-orbit splitting for excitons at the \textbf{K}(\textbf{K}') points in the Brillouin zone, as well as signatures of excitonic resonances at the $\Gamma$-point. 

The properties of the excitons at the \textbf{K}-point were extensively studied in the framework of  tight-binding and Bethe-Salpeter equation (BSE) \cite{berghauser2014analytical,Wu2015}, DFT+GW+BSE \cite{molina2013effect,qiu2013optical}, and gapped 2D Dirac-equation \cite{Wu2015}. 
One of the most prominent features
in the optical spectra of these materials is its dependence on the Berry phase, which  generates a modified Rydberg series \cite{zhou2015berry,srivastava2015signatures}. 
The form of the electron-electron interaction potential, which deviates form the Coulomb one, also contributes 
to a modified Rydberg series \cite{TonyRydberg}. 
The C-excitonic resonance in monolayer MoS$_2$, due to transitions at the $\Gamma$-point, 
was first calculated by Qiu \textit{et al.} \cite{qiu2013optical}, and it was associated with a minimum in the optical band structure around the
$\Gamma$-point by Klots \textit{et al.} \cite{MoS2exp1}. The effects of temperature and carrier density in MoS$_2$ were studied either  solving the semiconductor
Bloch equation (SBE) with a tight-binding Hamiltonian, whose parameters were obtained from a $G_0W$ calculation \cite{steinhoff2014influence}, or by  combining a LDA+BSE approach with the inclusion of electron-phonon coupling \cite{molina2016temperature}.

In the present work we use the polarization concept formalism \cite{haug2004quantum,Peres2010excitonic} for describing the 
excitonic properties of monolayer TMDC's. 
This formalism is easily applied to any system, both using low-energy effective models or tight-binding ones. The development of the formalism boils down to the solution of an eigenvalue problem for determining the excitonic bound states and to the solution of a linear system of equations for computing the optical conductivity of the system.
We apply the resulting equations to a two-band gapped Dirac equation
for describing the physics around the \textbf{K}-point; this  originates the physics of the A and B excitons in the TMDC's and of a modified Rydberg series. 
On the other hand, using as a starting point the three-band model for TMDC's \cite{Liu2013threeband} we describe the formation of an exitonic resonance near the $\Gamma$-point. This approach allows us to make much analytical progress and clearly identify the origin of different bound-states and resonances in the absorption spectrum. In this regards, our approach is distinct from previous ones that consider the full band structure as a starting point. The advantage of our approach lies in the possibility of clearly identify the origin of the different peaks in the optical conductivity, or absorbance for the same matters, of TMDC's.

We show that the optical properties have a strong dependence on external parameters, namely,  temperature and  dielectric function of the environment. This dependence on external parameters opens the possibility of engineering at will nano-materials showing strong optical response in the spectral range from the IR to visible. The application of these systems to opto-electronics, including photo-detectors, will launch a new set of devices in this area. Another
 possibility that these 2D materials may provide is the engineering  Bose-Einstein condensation of excitons when a TMDC is put inside an optical
 cavity \cite{vasilevskiy2015exciton}.

The paper is organized as follows: in section \ref{formalism} we introduce the second quantized form of the Hamiltonian, which is composed of three pieces:
the non-interacting part, the light-matter interaction term, and the Coulomb interaction. Using this Hamiltonian the excitonic properties at the Dirac point are
worked out. In section \ref{linear} we derive the optical properties of four TMDC's around the \textbf{K}-point in the Brillouin zone. Using a three-band tight-binding
model we describe the excitonic properties of four TMDC's in section \ref{gamma}. The excitonic effects around the $\Gamma-$point in the Brillouin zone are actually
resonances, as they are above the continuum. In section \ref{results} the optical properties of four TMDC's are discussed in detail and compared with the existent 
experimental data for the absorption. We find a good agreement with the experimental data, although, since we do not include electron-phonon interaction,
the agreement is not quantitative. 
We also note that the experimental values for the absorption present discrepancies among different experiments. This led us to conclude that there is
a clear sample-dependence in the absorption measurements.
In two of the TMDC's we stress the absence of the B-excitonic series in the experimental data, which is a noticeable discrepancy with our theoretical calculations. This led us to believe that the experiments need to be repeated for encapsulated TMDC's in h-BN at low temperatures.
This approach screens away the effect of extrinsic disorder, and reduces the 
impact of phonons in the absorption spectrum, due to low temperatures.
Finally, in section \ref{conclusions} we provide a summary of the main conclusions of the paper. A set of appendices give details of the calculations.

\section{Formalism and $\mathbf{K}$-point excitons\label{formalism}}
\begin{figure}
	\centering
	\includegraphics[scale=0.23]{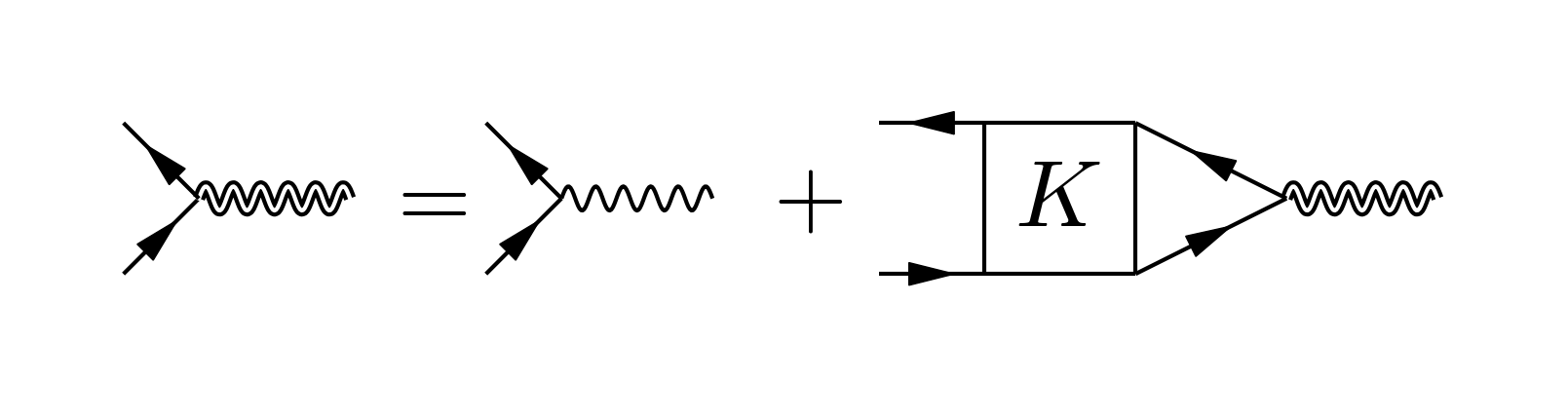}
	\caption{(Color on-line) Diagramatic expression of the Bethe-Salpeter equation for the vertex function. Our truncation of the  equation-of-motion for the density matrix (when we introduced the RPA approximation) is equivalent to consider the kernel of the BSE in the ladder approximation. However, a more precise approach would require more diagrams to be summed in the kernel.
	} \label{bse_diagram}
\end{figure}

In this section we introduce the effective model for electronic 
properties of TMDC's around the $\mathbf{K}$-point in the Brillouin zone. Since we are dealing with a many-body problem, the second quantization formalism is used throughout the paper. 
Using the full interacting Hamiltonian, the equations of motion for the density matrix are obtained and from it the total polarization is derived. The electron-electron interaction generates a hierarchy of
correlation functions that are 
truncated at the random-phase approximation (RPA) level. This procedure  is equivalent, in a diagrammatic approach, to the inclusion diagrams considering only the interaction in the electron-hole propagator. The diagrams relevant to our calculation 
are given in figure \ref{bse_diagram}, where $K$ represents the BSE
kernel (see ahead).
Although the diagrammatic approach is a possible route to solve the 
problem of excitonic effects in TMDC's, it is also possible to address it using an equation-of-motion approach. 
The latter  formalism enables treating at the same level of approximation the exchange-energy correction and the excitonic effects, and that is the 
path we will follow in this paper.

\subsection{Many-body Hamiltonian}

The low-energy single-particle electronic-excitations of TMDC materials can be described, in the $\mathbf{k}\cdot\mathbf{p}$
approximation, by a 2D gapped Dirac equation. When we consider spin-orbit coupling (SOC), the
effective mass and chemical potential become valley and spin dependent. With these aspects in mind, we
can write the single-particle Hamiltonian for a single combination of valley($\tau$)/spin($s$) index as:
\be
H_0^{s\tau}(\mathbf{k})=\hbar v_F\left(\tau \sigma_1 k_x+ \sigma_2 k_y \right)+\sigma_3 m_{s\tau} v_F^2-\mu_{s\tau} I, \label{H0_mos2}
\ee
with $\sigma_i$ the usual Pauli matrices and $I$ the identity matrix, $\tau=\pm$ the valley index, and $s=\pm$ the spin index. The effective mass, $m_{s\tau}$, and the on-site energy, $\mu_{s\tau}$, can be written in terms of the
SOC parameters, $\Lambda_1$ and $\Lambda_2 $ \cite{Kormanyos2015}, and of the mass
$\Delta$ as:
\be
m_{s\tau}=\Delta-\frac{s\tau}{2}\frac{\Lambda_1}{v_F^2},
\ee
\be
\mu_{s\tau}=\frac{1}{2}s\tau\Lambda_2\,.
\label{lambda2}
\ee
with $\Lambda_1=\Delta_{\rm{VB}}-\Delta_{\rm{CB}}$ and $\Lambda_2=\Delta_{\rm{VB}}+\Delta_{\rm{CB}}$, with
$\Delta_{\rm{VB}}$ ($\Delta_{\rm{CB}}$) the spin-splitting  of the valence (conduction) band.
\begin{figure}
	\centering
	\includegraphics[scale=0.53]{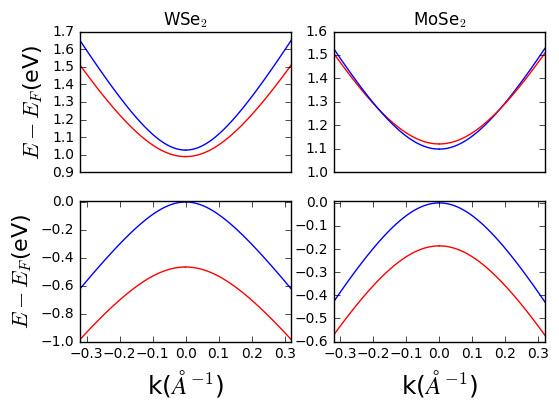}
	\caption{(Color on-line) Band structure of WSe$_2$ and MoSe$_2$ around the $\mathbf{K}-$point in the Brillouin zone, as described by Hamiltonian (\ref{H0_mos2}).
		Note that for WSe$_2$ the bands of different spin projections (different colors) do not cross, whereas for MoSe$_2$ there is a crossing
		in the conduction bands.  Due to these differences, the optical response of the two materials as function of doping differs from each other.
		In particular, in WSe$_2$ the highest-energy exciton peak is suppressed upon doping when compared to the lowest one.  (We have located the chemical potential at the top of the valence band.)
	} \label{fig:spin_bands}
\end{figure}
The band structure  implied by Hamiltonian (\ref{H0_mos2})  around the $\mathbf{K}-$point in the Brillouin zone is depicted in figure \ref{fig:spin_bands} (different colors
correspond to opposite spin projections).

In three dimensions (3D), the electron-electron interaction in a dielectric medium is given by the Coulomb potential in vacuum but with the 
permittivity of free space $\varepsilon_0$ 
replaced by the medium permittivity $\varepsilon_m\varepsilon_0$. Contrary to 3D, the same procedure does not hold in 2D materials.
In contrast, the electron-electron interaction is described by the Keldysh potential \cite{Cudazzo2011,rodin2014excitons}. This takes into account the surface charge polarization from a dielectric thin film
and reads in momentum space:
\be
V(q)=-\frac{e}{2 \varepsilon_0}\frac{1}{q(r_0q+\varepsilon_m)}, \label{Keldysh}
\ee
where 
$q$ is the 2D transfered momentum,
$\varepsilon_m$ and $r_0$ are the capping dielectric function of the environment 
and a
material-dependent constant, respectively, the latter measuring the deviation from the 2D Coulomb potential. Note that we  recover the 2D Coulomb potential making $r_0\rightarrow 0$. The potential (\ref{Keldysh}) is written in a slightly different manner than in
reference \cite{rodin2014excitons} for removing the dependence of the parameter $r_0$
on the external dielectric constant.

 To calculate the optical properties of  TMDC's, we consider the interaction of the electron gas with
a time-dependent electric field ${\cal E}(t)$ polarized along the $x$ axis.
For  describing  the light-matter interaction in this problem we use the dipole-coupling Hamiltonian
\be
\hat{H}_I(t)=-e{\cal E}(t) \hat{x},
\ee
with $\hat{x}$ the position operator.

From now on, we consider the full many-body Hamiltonian as:
\be
\hat{H}=\hat{H}_0+\hat{H}_I(t)+\hat{H}_\mathrm{ee}, \label{full_H}
\ee
where $\hat{H}_0$ is built from the single-particle Hamiltonian (\ref{H0_mos2}) and  $\hat{H}_\mathrm{ee}$ is the electron-electron interaction:  
\be
\hat H_\mathrm{ee}=  -\frac{e}{2}\int d\mathbf{r}_1 d\mathbf{r}_2\hat\psi^\dagger(\mathbf{r}_1) \hat\psi^\dagger(\mathbf{r}_2) V(\mathbf{r}_1-\mathbf{r}_2)\hat\psi(\mathbf{r}_2)\hat\psi(\mathbf{r}_1) , \label{hee_def}
\ee
where $\hat\psi(\mathbf{r})$ is the field operator (\ref{field_op}) 
defined below and $V(\mathbf{q})$ is the Fourier transform of the Keldysh potential given by equation (\ref{Keldysh}). For simplicity, from here on we choose units such that $v_F=\hbar=e=1$; the usual units are reintroduced at the end of the 
calculations. The field operator is given by:
\be
\hat\psi(\mathbf{r},t)=\frac{1}{L}\sum_{\mathbf{k},\lambda,s,\tau} \phi^{s\tau}_{ \lambda}(\mathbf{k})\hat{a}_{\mathbf{k}\lambda s \tau}(t)\e^{-\im\mathbf{k}\cdot\mathbf{r}}, \label{field_op}
\ee
with $L$ the square-box side length, $\hat{a}_{\mathbf{k}\lambda s \tau}$ the usual annihilation operator that obeys anticommutation relations and $\phi_{\mathbf{k}\lambda s \tau}$ the eigenfunctions of $H_0^\mathrm{s\tau}$, $
H_0^\mathrm{s\tau}\phi^{s\lambda}_{ \tau}(\mathbf{k})=(\lambda E_{k }^{s \tau}+\mu_{s\tau})\phi^{s\lambda}_{ \tau}(\mathbf{k}),
$
with $\lambda=-$ ($+$) for the valence (conduction) band. The eigenfunctions and positive eigenvalues are given by:
\be
\phi^{s\tau}_{ \lambda}(\mathbf{k})=\sqrt{\frac{E^{s\tau}_k+\lambda m_{s\tau}}{2E^{s\tau}_k}} \left(\begin{array}{cc} 1 \\ \frac{\tau k_x-\im k_y}{\lambda E^{s\tau}_k+m_{s\tau}} \end{array} \right), \label{eig_h0_ee}
\ee
\be
E^{s\tau}_{k}=\sqrt{k^2+m_{s\tau}^2}.
\ee
The eigenfunctions (\ref{eig_h0_ee}) will be used extensively in this work
for determining the four-body structure factor.

\subsection{Polarization operator}

For obtaining the optical conductivity and the absorbance, we have to compute the expectation value of the 
polarization operator. As noted above, we consider an electric field along the $x-$axis direction and 
define the polarization operator along the same spatial orientation. Using the  field operators  (\ref{field_op}) the polarization operator reads:
\be
\hat{P}(t)=\int d\mathbf{r}\,\,  \hat\psi^\dagger(\mathbf{r},t)(-ex)\hat\psi(\mathbf{r},t),
\ee
The integral can be explicitly computed with the help of the eigenfunctions of $H_0$ in  position space: $ \phi^{s\tau}_{\mathbf{k}\lambda}(\mathbf{r})= \phi^{s\tau}_{\lambda} (\mathbf{k})\e^{\im\mathbf{k}\cdot\mathbf{r}}$. Using these eigenfunctions it follows that
\bea
\int d\mathbf{r}\phi^\dagger_{\mathbf{k}^\prime\lambda^\prime s^\prime \tau^\prime} (\mathbf{r})  &x \phi_{\sigma\mathbf{k},\lambda}(\mathbf{r}) =
\left\langle s^\prime \tau^\prime\mathbf{k}^\prime,\lambda^\prime\left|x\right|s \tau\mathbf{k},\lambda\right\rangle 
=\nonumber\\&=\frac{\left\langle s^\prime \tau^\prime\mathbf{k}^\prime,\lambda^\prime\left|[x,H_0(x)]\right|s \tau\mathbf{k},\lambda\right\rangle}{\lambda E_{k}^{s\tau}-\lambda^\prime E_{k^\prime }^{s^\prime\tau^\prime}}\,.
\eea
Noting that $[x,H_0(x)]= -\im\sigma_1 $ we obtain for the dipole matrix element the result:
\be
\left\langle s^\prime \tau^\prime\mathbf{k}^\prime,\lambda^\prime\left|x\right|s\tau \mathbf{k},\lambda\right\rangle= \delta_{\mathbf{k}\mathbf{k^\prime}} \delta_{ss^\prime}\delta_{\tau\tau^\prime}\frac{\im v^{s\tau}_{\lambda^\prime}(\mathbf{k})}{ 2\lambda^\prime E_{k}^{s\tau}},
\ee
for $\lambda\ne \lambda^\prime$ (inter-band transitions). We defined the matrix element of the velocity operator $\sigma_1$ as
\be
v^{s\tau}_{\lambda}(\mathbf{k})=\left\langle s\tau\mathbf{k},\lambda\left|\sigma_1\right|s\tau\mathbf{k},-\lambda\right\rangle\,. \label{velocity_matrix}
\ee
Finally, we can express the polarization operator as:
\be
\hat P(t)= -\frac{\im e}{2}\sum_{s\tau\mathbf{k}\lambda} \frac{v^{s\tau}_{\lambda}(\mathbf{k})}{\lambda E_{k}^{s \tau}} \hat{\rho}^{s\tau}_{\mathbf{k}\lambda,-\lambda}(t),\label{Phat}
\ee
where we have introduced the following matrix in the Heisenberg picture: $\hat{\rho}^{s\tau}_{\mathbf{k}\lambda \lambda^\prime}(t)= \hat{a}^\dagger_{\mathbf{k},\lambda s\tau}(t) \hat{a}_{\mathbf{k},\lambda^\prime s\tau}(t)$. This matrix  is written in the basis that diagonalizes $\hat H_0$ and is equivalent to the density matrix for the states of $\hat H_0$. To determine the expectation value of the density matrix, and therefore the polarization, we use
the Heisenberg's equation-of-motion for $\hat{\rho}^{s\tau}_{\mathbf{k}\lambda \lambda^\prime}(t)$.

\subsection{Equations of motion for the density matrix}

As noted, for calculating the expectation value of the density matrix in the right-hand-side of equation (\ref{Phat}), we use  Heisenberg's equation-of-motion method. We define the expectation value for the off-diagonal elements of the density matrix, corresponding to the transition probabilities $p_{\lambda}^{s\tau}(\mathbf{k},t)$, as:
\be
p_{\lambda}^{s\tau}(\mathbf{k},t)=\left\langle\hat{\rho}^{s\tau}_{\mathbf{k}\lambda, -\lambda}(t) \right\rangle\,.
\ee
The diagonal elements of the matrix correspond to a new electronic distribution $n_\lambda^{s\tau}(\mathbf{k},t)$  defined as:
\be
n_\lambda^{s\tau}(\mathbf{k},t)=\left \langle \hat{\rho}^{s\tau}_{\mathbf{k}\lambda \lambda}(t)\right\rangle\,.
\ee
Both the diagonal and off-diagonal matrix elements of the density matrix are time-dependent.
Explicitly,  Heisenberg's equation-of-motion for the density matrix
reads:
\bea
-\im\frac{d}{dt} \hat{\rho}^{s\tau}_{\lambda \lambda^\prime}(t)=[\hat H&,\hat{\rho}^{s\tau}_{\lambda \lambda^\prime}(t)]=[\hat H_0,\hat{\rho}^{s\tau}_{\lambda \lambda^\prime}(t)]+\nonumber\\&+[\hat H_\mathrm{ee},\hat{\rho}^{s\tau}_{\lambda \lambda^\prime}(t)]+[\hat H_I,\hat{\rho}^{s\tau}_{\lambda \lambda^\prime}(t)], \label{mov_eq}
\eea
where we split the full Hamiltonian into the three components 
introduced in  equation (\ref{full_H}). The commutators in right-hand-side of equation (\ref{mov_eq}) are explicitly calculated in
 \ref{commutators}. 
The resulting equations-of-motion for the expectation value of equation (\ref{mov_eq}) are:
\bea
-\im\partial_t p^{s\tau}_{\lambda}(\mathbf{k},t)= &\left(\tilde{\omega}^{s\tau}_{\lambda\mathbf{k}}+{\cal W}_{\mathbf{k}\lambda}^{s\tau}(t) \right)  p^{s\tau}_{\lambda}(\mathbf{k},t)+\nonumber \\ &+ \big(\tilde{\Omega}^{s\tau}_{\mathbf{k}\lambda}(t)+{\cal D}^{s\tau}_{\mathbf{k}\lambda}(t) \big) \Delta n^{s\tau}_\lambda(\mathbf{k},t), \label{dbe_1}
\eea
\be
-\partial_t n^{s\tau}_\lambda(\mathbf{k},t)= 2\Im\left[ \left(\tilde{\Omega}^{s\tau}_{\mathbf{k}-\lambda}(t)+{\cal D}^{s\tau}_{\mathbf{k}\lambda}(t) \right) p^{s\tau}_\lambda\right], \label{dbe_2}
\ee \label{dirac_bloch_equation}
where the transition energy 
(also denoted bare optical band ahead) is $\omega_{\lambda\mathbf{k}}^{s\tau}=2\lambda E_\mathbf{k}^{s\tau}$, the difference in occupations reads $\Delta n^{s\tau}_\lambda(\mathbf{k},t)=n^{s\tau}_\lambda(\mathbf{k},t)-n^{s\tau}_{-\lambda}(\mathbf{k},t)$, and the renormalized Rabi frequency $\tilde{\Omega}^{s\tau}_{\mathbf{k}\lambda}(t)$ reads:
\be
\tilde{\Omega}^{s\tau}_{\mathbf{k}\lambda}(\mathbf{k},t)=-\frac{\im {\cal E}(t) v^{s\tau}_{-\lambda}(\mathbf{k})}{2\lambda E^{s\tau}_{\lambda k}}
+{\cal B}^{s\tau}_{\mathbf{k}\lambda}(\mathbf{k},t),
\ee
and depends on the dipole moment, the electric field, and the transition energy. The different terms
that appear in equations (\ref{dbe_1}) and (\ref{excitonic_rabi}) are classified as:
\begin{itemize}
\item Excitonic Rabi frequency  renormalization:
\bea
{\cal B}^{s\tau}_{\mathbf{k}\lambda}(t)= \frac{1}{S}\sum_{\mathbf{q}} V(|\mathbf{k}-\mathbf{q}|)\Big[p^{s\tau}_{\lambda}(\mathbf{q},t) F^{s\tau}_{\lambda^\prime\lambda^\prime\lambda\lambda}(\mathbf{k},\mathbf{q})+\nonumber\\
+ p^{s\tau}_{\lambda^\prime}(\mathbf{q},t) F^{s\tau}_{\lambda^\prime \lambda \lambda^\prime\lambda}(\mathbf{k},\mathbf{q})\Big], \label{excitonic_rabi}
\eea
and where in this expression we have the constraint $\lambda^\prime=-\lambda$, $S=L^2$ is the area of the system, and $F^{s\tau}_{\lambda_1,\lambda_2,\lambda_3,\lambda_4}(\mathbf{k_1},\mathbf{k_2})$
is defined in equation (\ref{F_definition})
\item Renormalized transition energy (or interacting optical band):
\be
\tilde{\omega}^{s\tau}_{\lambda\mathbf{k}}=2\lambda E^{s\tau}_k+\lambda\Sigma^{s\tau,\mathrm{xc}}_{\mathbf{k},\lambda},
\ee
where the exchange self-energy is:
\bea
\Sigma^{s\tau,\mathrm{xc}}_{\mathbf{k},\lambda}(t)=\frac{\lambda}{S}\sum_\mathbf{q} V(q) \Delta n^{s\tau}_\lambda(\mathbf{k}-\mathbf{q},t)\times\nonumber\\ 
\times\big[F^{s\tau}_{\lambda^\prime\lambda \lambda \lambda^\prime}(\mathbf{k},\mathbf{k}-\mathbf{q})
-F^{s\tau}_{\lambda\lambda\lambda\lambda}(\mathbf{k},\mathbf{k}-\mathbf{q}) \big], \label{exchange_self_energy_0}
\eea
and where in this expression we have the constraint $\lambda^\prime=-\lambda$. This form of the exchange self-energy is the same we find in the jelium model,
except for the non-trivial four-body structure factor  $F^{s\tau}_{\lambda_1\lambda_2 \lambda_3 \lambda_4}(\mathbf{k},\mathbf{k}-\mathbf{q})$.

\item Non-linear contribution:
\bea
{\cal W}^{s\tau}_{\mathbf{k}\lambda}(t)&=\frac{1}{S} \sum_{\mathbf{q} \lambda_1} V(|\mathbf{k}-\mathbf{q}|) p^{s\tau}_{\lambda_1}(\mathbf{q},t) \times\nonumber\\ &\times\big(F^{s\tau}_{\lambda^\prime \lambda_1^\prime\lambda_1\lambda^\prime}(\mathbf{k},\mathbf{q}) -F^{s\tau}_{\lambda \lambda_1^\prime\lambda_1\lambda}(\mathbf{k},\mathbf{q}) \big), \label{non_linear}
\eea
where in this expression $\lambda^\prime=-\lambda$, $\lambda_1^\prime=-\lambda_1$.

\item Density term:
\bea
{\cal D}^{s\tau}_{\mathbf{k}\lambda}(t)=\frac{1}{S} \sum_{\mathbf{q}\lambda_1} V(q)& n^{s\tau}_{\lambda_1}(\mathbf{k}-\mathbf{q},t) \times \nonumber \\ &\times F^{s\tau}_{\lambda^\prime\lambda_1\lambda_1\lambda} (\mathbf{k},\mathbf{k}-\mathbf{q}). \label{density}
\eea
\end{itemize}
Note that above we have defined the four-body spinor product (or structure factor) as:
\bea
F^{s\tau}_{\lambda_1,\lambda_2,\lambda_3,\lambda_4}(\mathbf{k_1},&\mathbf{k_2})=\nonumber \\ &{{\phi^{s\tau}_{\lambda_1}}^\dagger}(\mathbf{k}_1)
\phi^{s\tau}_{\lambda_2}(\mathbf{k_2})
{{\phi^{s\tau}_{\lambda_3}}^\dagger}(\mathbf{k_2})
\phi^{s\tau}_{\lambda_4}(\mathbf{k}_1), \label{F_definition}
\eea
that follows from writing the electron-electron interaction (\ref{hee_def}) in the basis that diagonalizes $H_0$ (\ref{eig_h0_ee}). Explicit equations for the above terms  are given in \ref{auxiliar_functions}. In the next sections we neglect the density $ {\cal D}^{s\tau}_{\mathbf{k}\lambda}(t)$ and the non-linear ${\cal W}^{s\tau}_{\mathbf{k}\lambda}(t)$ contributions, as  we are essentially focused on the exchange self-energy and  excitonic effects to linear order. We 
also consider that the system is in thermodynamic equilibrium, where the electronic distribution $n^{s\tau}_\lambda( \mathbf{k},t)$ is given by
the Fermi-Dirac distribution function $f^{s\tau}_\lambda(\mathbf{k})$:
\bea
n^{s\tau}_\lambda( \mathbf{k},t)=f^{s\tau}_\lambda&(\mathbf{k})=\nonumber\\
& \left[1+\exp\left(\frac{\lambda E^{s\tau}_\mathbf{k}-\mu_{s\tau}-E_F}{k_B T}\right)\right]^{-1}\,. \label{fermi_dirac}
\eea
The latter approximation is valid in the linear regime (weak external electric fields).

\subsection{Calculation of the exchange energy}

The exchange self-energy (\ref{exchange_self_energy_0}) reshapes the electronic bands  and, as will be shown later, it is essential to correctly describe the optical properties of TMDC's, specially the value of the independent-particle energy gap. 

 For graphene, described by a massless Dirac equation ($m^{s\tau}=0$), the exchange self-energy was calculated in \cite{Peres2005} and \cite{Hwang2007} with the use of the Couloumb potential. Here we calculate the exchange self-energy using the Keldysh interaction for the gapped Dirac equation ($m\ne0$).

The self-energy for the optical band structure, $\Sigma^{s\tau,\mathrm{xc}}(\mathbf{k})$, is calculated from equation (\ref{exchange_self_energy_0}) using the expressions in \ref{auxiliar_functions}. Its calculation boils down to an integral over all possible momentum
values:
\be
\Sigma^{s\tau,\mathrm{xc}}(\mathbf{k})=-\int \frac{d\mathbf{q}}{4\pi^2} V(q) \Delta f^{s\tau}_{\mathbf{k}-\mathbf{q}}\frac{\mathbf{k}\cdot\mathbf{q}+m_{s\tau}^2}{E^{s\tau}_k E^{s\tau}_q}, \label{self_energy}
\ee
where the difference between the electronic valence and conduction distribution functions is defined as $
\Delta f^{s\tau}_\mathbf{q}=f^{s\tau}_+(\mathbf{q})-f^{s\tau}_-(\mathbf{q}).
$

The direct gap renormalization, for each pair spin/valley, is given by the difference of the spin/valley top valence band and bottom conduction band energies, that is: 
\be
\Delta_{s\tau}=2m_{s\tau}+\Sigma^{s\tau,\mathrm{xc}}(\mathbf{k}=0), \label{gap_tmd}
\ee 
and for $T=0$, can be  calculated analytically from equation (\ref{self_energy}), resulting in
\bea
\Sigma^{s\tau,\mathrm{xc}}(\mathbf{k}=0)= \frac{ \alpha  m_{s\tau}}{\varepsilon_m \beta }\frac{Q(r_0 m_{s\tau}, k^{s\tau}_F/m_{s\tau})}{\sqrt{1+(r_0m_{s\tau})^2}}, \\
Q(\zeta,\xi)=\ln\left[\frac{\zeta\left(\zeta-\xi+\sqrt{\xi^2+1}\sqrt{\zeta^2+1}\right)}{(\zeta \xi+1)(\sqrt{1+\zeta^2}-1)} \right],\label{gap_analytic}
\eea
with $\alpha\approx1/137$ the fine structure constant, $\beta=v_F/c$ the ratio between Fermi-velocity and the speed of light, and the valley-spin
dependent Fermi momentum $k_F^{s\tau}$ is given by:
\be
k_F^{s\tau}=\sqrt{(E_F+\mu_{s\tau})^2-m_{s\tau}^2}.
\ee 

\begin{figure}  
   \vspace{0.5cm}
   \centering
   \includegraphics[scale=0.5]{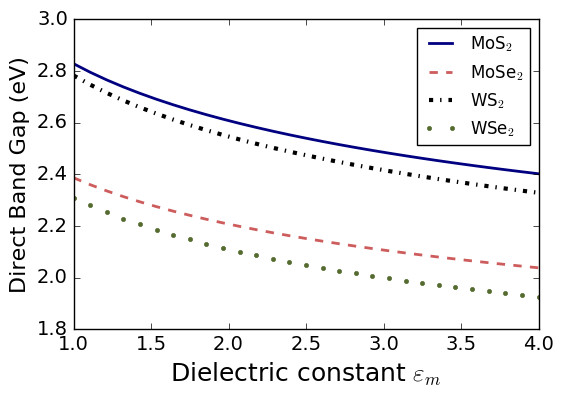}
   \caption{ (Color on-line) Dependence of the bandgap renormalization in TMDC's (\ref{gap_analytic}), computed from the exchange self-energy, on the capping dielectric function at $T=0$ K.
   	A higher dielectric constant suppresses the electron-electron interactions and thus the renormalization of the band gap by exchange energy becomes smaller. 
   	The parameters used are from table \ref{parameters_table}.} \label{gap_ren}
\end{figure}

\begin{figure}  
   \vspace{0.5cm}
   \centering
   \includegraphics[scale=0.5]{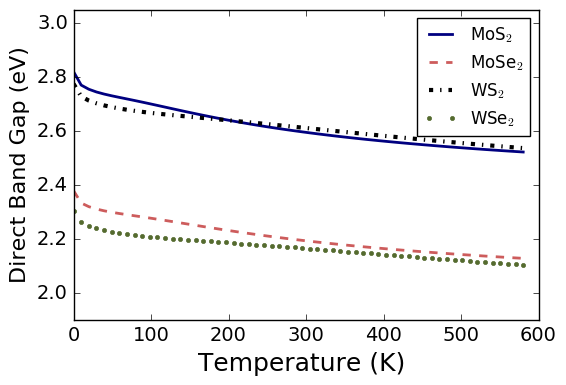}
   \caption{(Color on-line) Temperature dependence of the bandgap renormalization in TMDC's
   	computed 
   	 from the exchange self-energy (\ref{self_energy}). Parameters 
used are from table \ref{parameters_table}.The chemical potential is set at the top of the valence band, but any other location would give 
qualitatively similar results. 
 As the temperature increases, for a fixed chemical potential, the valence band depopulates,  less carriers are available in the band and, as a consequence,
  the exchange self-energy decreases. } \label{tep_ren}
\end{figure}

\begin{figure}  
   \vspace{0.5cm}
   \centering
   \includegraphics[scale=0.4]{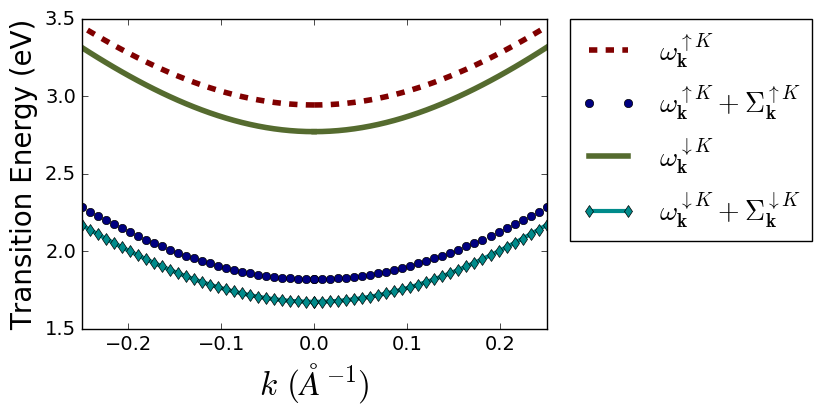}
   \caption{(Color on-line) 
   	Renormalization of the
   	transition energy $\omega_\mathbf{k}=2E_\mathbf{k}$  in MoS$_2$ due to
   	 the exchange self-energy (\ref{self_energy}). We can see the band gap shift and an increase in the curvature of the band, relatively to
   	 the independent particle approach. Parameters used
were $m=0.796$ eV and  $q_0=1/33$ \r{A}$^{-1}$ \label{band_ren} }
\end{figure}

We depict  the
dependence of gap-renormalization 
 in figure (\ref{gap_ren}).  The temperature dependence of the same quantity is given in figure (\ref{tep_ren}), and the renormalized band in figure (\ref{band_ren}). We can see the strong dependence of the exchange energy on the external parameters (temperature and dielectric constant of the medium surrounding the TMDC), showing that the environment plays a key role on the optical properties of these materials.

\begin{table*}
\begin{tabular}{c|c|c|c|c|c|c|cccccc}
 TMDC &  $\Delta $(eV)  &  $\hbar v_F$(eV/\AA) & $\Lambda_1$(eV)  & $\Lambda_2$(eV)  &  $r_0$(\AA)  & $\Delta_K$(eV) & &  Experimental gap\\ \hline
MoS$_2$  & 0.797 & 2.76 & 0.076 & 0.073 & 31.4 & 2.82 & 2.5 \cite{MoS2exp1}  & 2.14 \cite{MoS2exp2} & 1.86 \cite{MoS2exp3}\\
MoSe$_2$ & 0.648 & 2.53 & 0.104 & 0.082 & 51.7 & 2.37 & 2.18 \cite{MoSe2exp1} & 2.02, 2.22 \cite{MoSe2exp2}  & 1.58 \cite{MoSe2exp3}\\
WS$_2$   & 0.685 & 3.34 & 0.164 & 0.230 & 37.9 & 2.78 & 2.14 \cite{MoS2exp3} & 2.41 \cite{WS2exp2}\\
WSe$_2$  & 0.524 & 3.17 & 0.215 & 0.252 & 45.1 & 2.31 & 2.51 \cite{WSe2exp1} & 2.0, 2.18 \cite{MoSe2exp2}\\
\hline
\end{tabular}
\caption{The first five columns give the material parameters used in all calculations in this paper. 
	 The parameters in the first four columns come
	from Ref. \cite{Kormanyos2015} and in the fifth column from Ref. \cite{Berkelbach2013}.
	The sixth column is the direct gap at the \textbf{K}-point calculated with
the exchange self-energy obtained from our model. The last columns are experimental data (numbers in square braces refer to references).
See also \cite{Rossier2013}	for a different set of parameters and, in particular, the prediction of SOC induced splitting in the conduction band at the
\textbf{K}-point.}
 \label{parameters_table}
\end{table*}

A more accurate approach to the calculation of renormalization of the band gap  due to  electron-electron interactions requires a self-consistent approach, where the unperturbed Hamiltonian $H_0^{s\tau}$ (\ref{H0_mos2}) is defined 
including the self-energy from electron-electron interactions. For graphene this procedure
is used to study the possibility of  dynamical  generation of a gap in the spectrum \cite{Kotov2012,Chaves2011}.

\subsection{Excitonic effects in the Rabi frequency}

The electron-electron interaction induces the creation of electron-hole bound states below the non-interacting energy gap  (corrected by the exchange self-energy), that corresponds
to excitonic bound states. Also excitonic resonances appear above the energy gap. These two effects are routinely measured in optical experiments in
semiconductors \cite{haug2004quantum}.
In the equation-of-motion description, the electron-electron
interaction renormalizes the Rabi frequency, and, as will be shown later, this corresponds to solve the Bethe-Salpeter equation in the ladder approximation and in the center-of-mass reference frame. This procedure allows  the calculation of the renormalized  expectation value of the $\hat{x}$ operator.

The renormalized Rabi frequency (\ref{excitonic_rabi}), term ${\cal B}^{s\tau}_{\mathbf{k}\lambda}(t)$, can be split into two parts (addition of two terms):
\bea
{\cal B}^{s\tau}_{\mathbf{k}\lambda}(t)
=\frac{1}{S}\sum_{\mathbf{q}} V(|\mathbf{k}-\mathbf{q}|)p^{s\tau}_{\lambda}(\mathbf{q},t) F^{s\tau}_{-\lambda-\lambda\lambda\lambda}(\mathbf{k},\mathbf{q})+\nonumber\\
+\frac{1}{S}\sum_{\mathbf{q}} V(|\mathbf{k}-\mathbf{q}|)p^{s\tau}_{-\lambda}(\mathbf{q},t) F^{s\tau}_{-\lambda\lambda-\lambda\lambda}(\mathbf{k},\mathbf{q})\,. \label{rabi_frequency_2}
\eea
When $\lambda=+$ the first term corresponds to the
generation of an electron-hole pair before  scattering
from the Coulomb potential. The second term is not included in the usual approach to the  Wannier equation, but has a non-negligible contribution to the optical conductivity.


\section{Approximated equation-of-motion for the polarization operator: Linear response \label{linear}}

In this section we derive the frequency response, in the linear regime, for equation (\ref{dbe_1}) using the thermodynamic equilibrium density function given
by the Fermi-Dirac distribution (\ref{fermi_dirac}). The external electrical field is written as ${\cal E}(t)={\cal E}_0\e^{\im\omega t} $,
and the linear-response is obtained from the terms proportional to $\e^{\im\omega t} $, $p^{s\tau}_\lambda(\mathbf{k},t)=p^{s\tau}_\lambda(\mathbf{k},\omega)e^{\im\omega t} $. In this regime we neglect the non-linear term ${\cal W}^{s\tau}_{\mathbf{k}\lambda}(t)$ (\ref{non_linear}) that only contributes to second-harmonic
and higher frequency terms, and the density term ${\cal D}^{s\tau}_{\mathbf{k}\lambda}(t) $, that contributes indirectly to the linear response through  equation (\ref{dbe_2}). With these considerations, the equation-of-motion (\ref{dbe_1}) becomes:
\be 
\left(\omega -\tilde{\omega}^{s\tau}_{\lambda\mathbf{k}} \right) p^{s\tau}_{\lambda}(\mathbf{k},\omega)= \left( \frac{{\cal E}_0 v^{s\tau}_{-\lambda}(\mathbf{k})}{2\im\lambda E^{s\tau}_k}
+{\cal B}^{s\tau}_{\mathbf{k}\lambda}(\omega) \right) \Delta f^{s\tau}_\mathbf{k},\label{ap_eq_mot}
\ee 
with ${\cal B}^{s\tau}_{\mathbf{k}\lambda}(\omega) $ obtained replacing $p^{s\tau}_\lambda(\mathbf{k},t)$ by $p^{s\tau}_\lambda(\mathbf{k},\omega)$ in equation (\ref{rabi_frequency_2}).

Once equation (\ref{ap_eq_mot}) is solved for the transition probabilities $p^{s\tau}_\lambda(\mathbf{k},\omega)$, the total polarization $P(t)=P(\omega)e^{\im\omega t}$ can be obtained from the expectation value of equation (\ref{Phat}):
\be
P(\omega)= -S\sum_{s\tau\lambda} \int\frac{d\mathbf{k}}{(2\pi)^2} \frac{v^{s\tau}_{\lambda}(\mathbf{k})}{2\lambda E_{k}^{s\tau}} p^{s\tau}_{\lambda}(\mathbf{k},\omega), \label{pfinal}
\ee
and the optical conductivity follows from the macroscopic relation between the polarization current density
and the polarization density $\mathbf{J}(t)= \partial_t \mathbf{P}(t) $:
\be
\sigma(\omega)= \im\frac{\omega}{S {\cal E}_0}P(\omega). \label{sigma_P_relation}
\ee

 The presence of the excitonic term,  equation (\ref{rabi_frequency_2}) in ${\Omega^{1}}^{s\tau}_{\mathbf{k}\lambda}(\omega)$,
 makes the
equation of motion (\ref{ap_eq_mot}) a system of two coupled Fredholm integral equations of the second kind for the transitions
probabilities $p^{s\tau}_{\lambda}(\mathbf{k},\omega)$, $\lambda=\pm$, that has to be solved for each spin/valley pair $s\tau$. The correspondent
homogeneous equation [that can be obtained making ${\cal E}_0=0$ in equation (\ref{ap_eq_mot})], corresponds to a Fredholm integral equation of the first kind.  The solution of the latter  will be explored in the next section.

\subsection{Study of homogeneous Bethe-Salpeter Equation: Excitonic states}

The set of equations (\ref{ap_eq_mot}), in the homogeneous case, corresponds to the Bethe-Salpeter equation for the
exciton wave function $\psi_{nl}^{s\tau}$ with energy $E^{s\tau,\mathrm{exc}}_{nl}$. In this limit this equation is also known as the Wannier equation \cite{haug2004quantum}. 
Once  the excitonic wave functions are known we can calculate the absorbance coefficient ${\cal A}(\omega)$
using  Elliot's formula \cite{haug2004quantum},  in a form appropriate for TMDC's;  this formula  is derived in \ref{elliot_appendix} following a procedure described in \cite{berghauser2014analytical}; this approach leads to 
\be
{\cal A}(\omega)\approx\frac{4\pi\alpha\omega\gamma}{\sqrt{\varepsilon_m}}  \sum_{s\tau, l=\{0,2\},n}  \frac{M^{s\tau}_{n\ell}}{(\omega-E^{s\tau}_{n\ell }/\hbar)^2+\gamma^2} ,
\ee
where $M^{s\tau}_{n\ell}$ is the oscillator strength, given by:
\be
M^{s\tau}_{n\ell}= v_F^2 \left|\int_0^\infty q dq \frac{v^{s\tau}_{\ell,+}(q)}{2 E^{s\tau}_q   } \left[\psi^{s\tau}_{n\ell}(q) \right]^* \right|^2, \label{weight_elliot}
\ee
with $\ell$ and $n$ the angular and radial quantum numbers, and we have explicitly reintroduced the Fermi velocity $v_F$ for defining
the oscillator strength as a dimensionless quantity.

In the system of equations (\ref{ap_eq_mot}) we neglect the non-resonant term $p^{s\tau}_{-}(\mathbf{k},\omega)$,  consider
the system at zero temperature, and at the charge neutrality point, $\Delta f^{s\tau}_\mathbf{k}=-1$.  Thus,  we have 
for the homogeneous problem an integral equation for $p^{s\tau}_{+}(\mathbf{k},\omega) $:
\bea
(\omega- \tilde{\omega}^{s\tau}_{+ k})p^{s\tau}_{+}(\mathbf{k},\omega)=\nonumber\\
-\int \frac{d\mathbf{q} }{(2\pi)^2}  V(|\mathbf{k}-\mathbf{q}|) F^{s\tau}_{++++}(\mathbf{k},\mathbf{q}) p^{s\tau}_{+}(\mathbf{q},\omega)\,. \label{bse_aprox}
\eea
If we also neglect the exchange self-energy term in the previous result, equation (\ref{bse_aprox}) is formally equal to the Bethe-Salpeter equation for the two-body electron-hole wave function obtained in the center-of-mass
reference frame in a gapped Dirac system. We can write equation (\ref{bse_aprox}) in the following matrix form:
\be
\left(E_\mathrm{exc}-K^{BS}\right) \Psi=0, \label{bse_matrix}
\ee
with $E_\mathrm{exc}=\omega$ the Exciton energy and $K^{BS}$ the integral operator of equation (\ref{bse_aprox}). We can use the cylindrical symmetry to write the eigenfunctions of equation (\ref{bse_aprox}) as:
\be
p^{s\tau}_{+}(k,\theta,\omega)=\sum_{n\ell} \psi^{s\tau}_{n\ell}(k)\e^{\im\theta} \e^{\im \ell\theta}, \label{def_eq}
\ee
with $\ell$ the angular quantum number (note that we have omitted the dependence of 
$\psi^{s\tau}_{n\ell}(k)$ in $\omega$). The extra phase in definition (\ref{def_eq}) allows to write equation (\ref{bse_aprox}) in the heavy mass limit $m\rightarrow\infty$ as a hydrogen-atom equation in momentum space
with a screened potential \cite{berkelbach2015}. The introduction of a factor $\e^{\im\theta}$ to classify  the excitons can be seen as arbitrary. Since the spinors, given by equation (\ref{eig_h0_ee}), also have an arbitrary global phase that propagates to  the four-spinor product (\ref{F_definition}),  the best way to define the $s$-wave is by a limit condition, that is, taking the  limit
$m\rightarrow\infty$ we should recover the spectrum of the hydrogen atom. 

Substituting (\ref{def_eq}) into (\ref{bse_aprox}), and using the  orthogonal relations for the wave function $\psi^{s\tau}_{n\ell}(k)$, equation (\ref{bse_aprox}) becomes the following Wannier formula:
\be
(\omega- \tilde{\omega}^{s\tau}_{+k})\psi^{s\tau}_{n\ell}(k)=\int_0^\infty dq  \tilde{T}_\ell^{s\tau}(k,q)\psi^{s\tau}_{n\ell}(q), \label{bse_redux}
\ee
with the kernel $\tilde{T}_\ell^{s\tau}(k,q)$ given in \ref{bethe_salpeter_kernel}. Equation (\ref{bse_redux}) was solved
before in references \cite{berghauser2014analytical,Wu2015} without the exchange self-energy contribution in $\tilde{\omega}^{s\tau}_{+k}$. Here we include the effect of the exchange term.

\begin{figure}
\centering
   \includegraphics[scale=0.6]{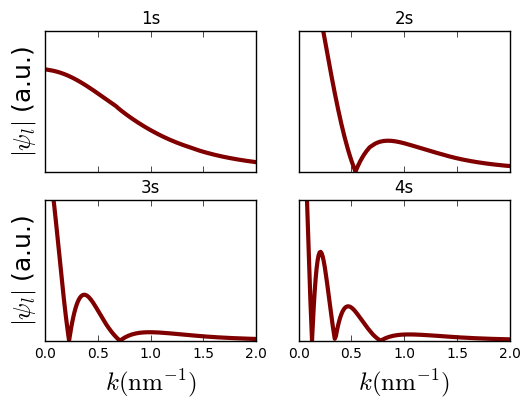}
   \caption{(Color on-line) Unnormalized radial excitonic wave function
   	for MoS$_2$ from the solution of equation
   	 (\ref{bse_redux}) for $\ell=0$ at the $\mathbf{K}-$point.  We can see the usual increase of nodes with the higher modes. 
   	 The size of the exciton, estimated as $2\pi/k$, is of the order of  $\sim10$ nm, or about  few of unit cells. } \label{swavefunction}
\end{figure}

The results for  convergence of the binding energies, $E^\mathrm{binding}_{n\ell}=\Delta_{s\tau}-E_{n\ell}$, with $\Delta_{s\tau}$ given by equation (\ref{gap_tmd}), of the integral equation for the $s$-wave ($\ell=0$) and the $p$-waves ($\ell=\pm 1$) 
are presented in columns labeled ``3000" and ``GL" of table \ref{bse_table} (in it we also discuss convergence issues of our numerical methods). In this table we compare our results with those of reference \cite{Wu2015}, and because of this we have neglected the exchange correction as those authors also did. The wave functions for the $s$-wave are presented in Fig. (\ref{swavefunction}), where
we can see the usual increase of nodes for higher modes. The breakdown of degeneracy in $p$-waves comes from the $\ell$ dependence in the matrix
element of the four body spinor product, that appears inside the integral  (\ref{int_func}), and is a consequence of the Berry curvature of the Dirac Hamiltonian \cite{zhou2015berry,srivastava2015signatures}.

\begin{table*}
\centering
\begin{tabular}{ccccccccc}
 &  50 &  100 & 300 &  600 & 1200 & {\bf 3000} & {\bf GL} & LM Ref.\cite{Wu2015}\\\hline
1s & 0.224 & 0.264 & 0.304 & 0.318 & 0.327 & 0.333 &0.331 & 0.301 \\
2s & 0.033 & 0.055 & 0.081 & 0.092 & 0.099 & 0.104 & 0.103 & 0.099\\
2p$^+$ &0.051 & 0.077 & 0.107& 0.119 & 0.126 & 0.132 & 0.132&0.125\\
2p$^-$ &0.062 & 0.090 & 0.122 & 0.134 & 0.142 & 0.148 &0.147 &0.150\\
\hline
\end{tabular}
\caption{
	Binding energy (in eV) for four different excitonic states of MoS$_2$ corresponding to the A-series.
	The first six data rows show the results for the numerical procedure based on a constant grid discretization; convergence is obtained only 
	for very large grids of the order of 3000. GL accounts for Gauss-Legendre/Laguerre, where
the integral in $q$ is divided in three intervals $[0,0.3\Delta]$, $[0.3\Delta,0.6\Delta]$, $[0.6\Delta,\infty]$. For the first interval we use
100 points, and for the second and third 50 points are used. In the first two intervals we use Gauss-Legendre quadrature, and the last one Gauss-Laguerre with
a rescale of $m/150$. The last row (LM) is the data from reference \cite{Wu2015}, with excitonic binding energy
computed using
a lattice (tight-binding) model. We want to stress that the discretization procedure in a linear mesh of 3000 points takes several hours to run in a conventional laptop, whereas the GL method takes only few minutes in the same computer.
At a given stage the calculation requires performing an angular integral of the Keldysh potential $V(\mathbf{k}-\mathbf{q})$. Although the integral can be proven finite,
numerically the integral is ill behaved when $k=q$. For avoiding this pathology a finite constant of the order of the 
grid spacing is added to $p_{k,q}(\theta)$ in Eq. (\ref{int_func}); this renders the integral finite, as it should.
}
 \label{bse_table}
\end{table*}

\begin{figure}
\centering
   \includegraphics[scale=0.6]{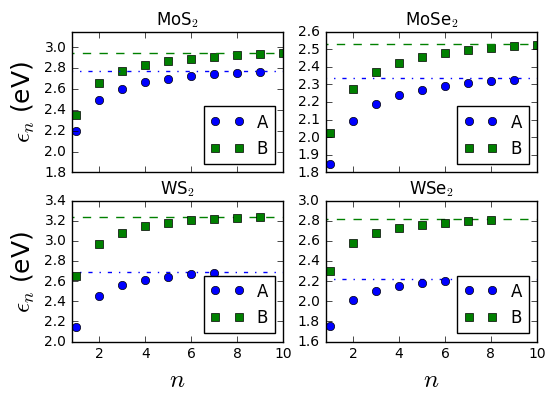}
   \caption{ (Color on-line) Exciton eigenvalues of equation (\ref{bse_matrix}), with exchange effects included, as function of the
   	radial quantum number $n$ for $\ell=0$. The dashed line corresponds to the spin/valley dependent gap, given by 
equation (\ref{gap_analytic}). The blue circles (green squares) corresponds to the $\uparrow K$, $\downarrow K^\prime$ ($\downarrow K$,
$\uparrow K^\prime$) series.
Note that for WS$_2$ and WSe$_2$ the  B excitons are all but one (1$s$=B) inside the continuum of the A excitonic series.
This fact is expected to have important consequences in the absorbance spectrum of these two materials (see section \ref{results}).
} \label{exciton_exchange_bse}
\end{figure}

The Rydberg series for excitons including exchange corrections is presented in figure \ref{exciton_exchange_bse} for all four TMDC's considered in this work with the parameters of table \ref{bse_table}. There are
four combinations of spin/valley, but the time-reversal symmetry reduces this number to two independent combinations. Each of these will generate a distinct Rydberg series, that we call A for the lowest fundamental energy and B for the highest one. Note that the A and B series are split due to SOC.
For WS$_2$ we compare the Rydberg series with the experimental work
of ref. \cite{WS2exp2};  our results show a good agreement with the experimental data, with a  deviation smaller than $5$ meV.

\begin{figure}
\centering
   \includegraphics[scale=0.6]{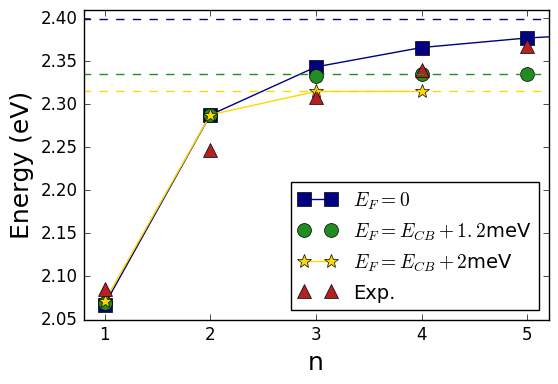}
   \caption{ (Color on-line) Comparison between experimental data (red triangles) from Ref. \cite{WS2exp2} and our theoretical model, considering a small amount of doping (green circles and yellow stars), and the neutral case (blue squares). Exciton eigenvalues of equation (\ref{bse_matrix}) for WS$_2$, with exchange effects included, as function of the
   	radial quantum number $n$ for $\ell=0$. The dashed line corresponds to the spin/valley dependent gap, given by 
equation (\ref{gap_analytic}). The parameters of the mass, Fermi velocity and SOC are from table \ref{bse_table}. We set $r_0=40.92\,\,$ \AA,
slightly larger than that given in reference \cite{Berkelbach2013}, $37.89\,\,$\AA.
The effective dielectric constant, including the effect of the substrate (SiO$_2$), is $\varepsilon=2.45$.
} \label{rydberg}
\end{figure}
\subsection{Integral equation for the vertex function}

With the knowledge of the solution of the homogeneous equation, we come back to the integral equation (\ref{ap_eq_mot}),
where we add a phenomenological interband relaxation rate $\gamma_p$ to include disorder effects. 

We write
$p^{s\tau}_{\lambda }(k,\theta,\omega)={\cal E}_0 \Psi^{s\tau}_{\lambda}(k,\theta)$ and proceed as
we did in equation (\ref{def_eq}) expanding $\Psi^{s\tau}_{\lambda}(k,\theta)$
in the eigenstates of the homogeneous Bethe-Salpeter equation and angular momentum states:
\be
\Psi^{s\tau}_{\lambda}(k,\theta)=\sum_{\ell=-\infty}^\infty \psi^{s\tau}_{\lambda\ell}(k) e^{\im\theta} e^{\im\ell\theta},
\ee
and as a consequence:
\bea
(\omega-&\tilde{\omega}^{s\tau}_{\lambda k}+i\gamma)  \psi^{s\tau}_{\lambda\ell} (k)=\Delta f^{s\tau}_k\Bigg[
\frac{ v^{s\tau}_{0,\lambda}(k)}{2\lambda E_k^{s\tau}}\delta_{\ell,0}+
\nonumber\\ &+ \frac{v^{s\tau}_{-2,\lambda}(k)  }{2\lambda E_k^{s\tau}}\delta_{\ell,-2}+\int_0^\infty dq \Big( T^{1,s\tau}_{\lambda,\ell} (k,q)\psi^{s\tau}_{\lambda \ell}(k)
+\nonumber \\  & +  T^{2,s\tau}_{\lambda,\ell}(k,q)  \psi^{s\tau}_{-\lambda\ell}(k)\Big)\Bigg], \label{ap_eq_vertex_1}
\eea
where $\delta_{\ell,s}$ is the Kronecker-delta, and the kernels $T^{1,s\tau}_{\lambda,\ell} (k,q)$ and  $T^{2,s\tau}_{\lambda,\ell} (k,q)$, and the
velocity angular decomposition $v^{s\tau}_{\pm,\lambda}(k)$ are given in \ref{bethe_salpeter_kernel}. 

The diagrammatic representation of equation (\ref{ap_eq_vertex_1}) is shown in figure \ref{bse_diagram}, where the internal electron-hole legs are understood to be dressed by the exchange interaction.

\subsection{Bright and Dark excitons: optical selection rules}

We are now in position to discuss the conditions that an exciton can absorb a photon. From the coupled set of equations (\ref{ap_eq_vertex_1}), only $\ell=0$ ($s$-excitons) and $\ell=-2$ ($d$-excitons) contribute to the optical conductivity, but from our numerical calculations, the $\ell=-2$ mode contribution is negligible for real TMDC's parameters.  After performing the  $\theta$-integration in equation (\ref{pfinal}), we obtain
for the polarization:
\be
\frac{P(\omega)}{S{\cal E}_0}= \sum_{s\tau\lambda} \int\frac{k dk}{2\pi} \frac{v^{s\tau}_{0,\lambda}(k) \psi^{s\tau}_{\lambda,-2}(k)+v^{s\tau}_{-2,\lambda}(k) \psi^{s\tau}_{\lambda,0}(k)}{-2\lambda E^{s\tau}_{k}}. \label{p_num}
\ee

Finally, the numerical procedure to calculate the optical conductivity is the following: we solve the integral equation (\ref{ap_eq_vertex_1}) to determine
the eigenfunctions  $\psi^{s\tau}_{\lambda\ell} (k)$, calculate the polarization from the integral in equation (\ref{p_num}), and
lastly the optical conductivity is calculated from  relation (\ref{sigma_P_relation}). The absorbance for a MoS$_2$ suspended sheet for a TEM wave with normal incidence can be obtained from
the optical conductivity as \cite{Goncalves2016}:
\be
{\cal A}(\omega)= \alpha\pi \frac{4\Re\left[f(\omega)\right]}{4+ \pi^2\alpha^2 |f(\omega)|^2},
\label{eq_absorb}
\ee
with $f(\omega)=\sigma(\omega)/\sigma_0$ and $\sigma_0=e^2/4\hbar$.


\section{Excitons at the $\Gamma$-point\label{gamma}}

To  describe accurately the optical absorption in the frequency domain after the two first excitonic peaks (A and B), that is the region roughly located in the interval  $2.3-3.5$ eV, we need
to describe the excitonic effects due to electronic transitions at the 
$\Gamma$-point.
This implies  going beyond the $\mathbf{k}\cdot\mathbf{p}$ model at $\mathbf{K}$-point. To accomplish this,
we use the three-band model of Liu {\it et al.} \cite{Liu2013threeband}, which was shown to 
describe accurately the GGA band structure. We use the same equation of motion approach introduced in
previous section. The dipole matrix element is calculated using a Peierls approximation \cite{Tomczak2009Peierls,steinhoff2014influence}:
\be
\langle \lambda \mathbf{k}| \hat x| \lambda^\prime \mathbf{k} \rangle = \frac{i}{E^{\lambda^\prime}_\mathbf{k}-E^\lambda_\mathbf{k}} \langle \lambda \mathbf{k}|\partial_\mathbf{k} H_\mathbf{k}| \lambda^\prime \mathbf{k} \rangle, \label{dip_gamma}
\ee
which takes in account vertical interband transitions only. 

\begin{figure*}  
   \vspace{0.5cm}
   \centering
   \includegraphics[scale=0.3]{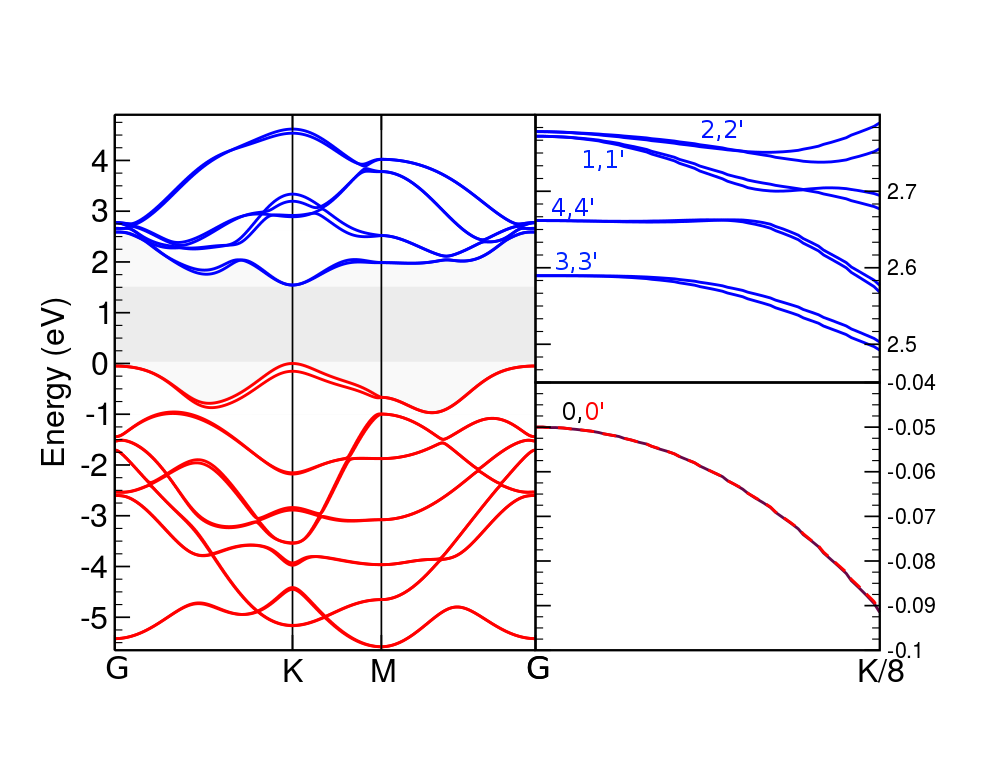}
   \caption{(Color on-line)  Electronic bands diagram of single layer MoS$_2$. In red are the valence bands and in blue the conduction bands. On the right panels a zoom-in of the conduction bands (top) and valence bands (bottom) near the $\Gamma$ point is shown.
The band diagrams were obtained using Density Functional Theory in the GGA (PBE)\cite{Perdew1996} approximation, as implemented in the {\sc Quantum ESPRESSO}\cite{Giannozzi2009} package. An energy cutoff of 70 Ry and a Monkhorst-Pack\cite{Monkhorst1976} grid of $16\times 16\times 1$ were used. Mo and S atoms are represented by norm conserving pseudopotentials generated with fully relativistic calculations including spin-orbit interaction. To avoid interaction between different images of the layer, a 45 bohr supercell in the $c$ direction is included.} \label{MoS2_band_structure}
\end{figure*}

To proceed with the discussion about excitons (actually excitonic resonances) 
at the $\Gamma$-point, we have to look
in detail into the TMDC band structure depicted in figure \ref{MoS2_band_structure}; this 
band-structure was calculated using a full relativistic method (see figure caption for details), that is necessary
to correctly account for the spin-orbit coupling \cite{carvalho2013band} in TMDC's. 
Very close to the $\Gamma$-point, the top of the valence band (that from now on we label band $0$) and the 
last four conduction bands (that, from here on, are labeled $1$, $1^\prime$, $2$ and $2^\prime$, in increasing energy order) are essentially due to contributions from electrons belonging to the transition-metal $d$-orbitals. We note in passing that the bands $0$, $1$, and $2$ are also used in the effective Hamiltonian valid near the $K$-point. The four lowest conduction bands at the $\Gamma$ point (labeled $3$, $3^\prime$, $4$ and $4^\prime$, in increasing energy order) are mostly composed of $p$-orbitals from the chalcogenides atoms. From here on, we do not consider SOC effects in the three band model as they are very small at the $\Gamma-$point (see right panels in Fig. \ref{MoS2_band_structure}). Therefore, we drop the prime notation of the bands, that is, the bands $i$ and $i^\prime$ are treated as spin-degenerated (at the computation level this amounts to a multiplicative factor of two affecting the
absorbance curves).

The mirror symmetry, with respect to the plane formed by the metallic atoms, is important to discuss the optical properties of TMDC's. The bands $0$, $1$ and $2$ all have even symmetry, while the bands $3$ and $4$ both have odd symmetry.  Therefore, the dipole matrix elements between bands with different mirror symmetry vanish (note that the mirror symmetry refers here to the $z-$coordinate). As a consequence, the existence of excitons composed of holes from the $0$ band and electrons from the bands $3$ and $4$, they are optically dark under a single-photon experiment. This implies that we restrict the calculation of excitonic effects near the $\Gamma-$point considering only the optical properties of excitons composed of holes from   band $0$ and electrons from bands $1$ and $2$. These optical transitions generate a modified  Rydberg series, which is a consequence of the non-parabolic dispersion relation of the {\it optical band} $[$see equation (\ref{gamma_obs})$]$ and finite Berry curvature at the $\Gamma$-point and as well as from the form of the Keldysh potential.

To calculate the optical properties of the excitons at the $\Gamma$-point, we use the same equation of motion method developed in previous section. We define  transition probabilities $p_{0i}(\mathbf{k},t)=\langle \hat{a}^\dagger_{i\mathbf{k}}(t) \hat{a}_{0\mathbf{k}}(t)\rangle$, that represent the annihilation of an electron at the valence band $0$ and a creation of an electron at the conduction band $i=1,2$. The equation of motion
is again given by (\ref{mov_eq}), making $\lambda=0$ and $\lambda^\prime=1,2$, the latter two values represent the two possible bright excitons at the $\Gamma$-point. On the other hand, the  $H_0=H_0^{3\mathrm{b}}$ Hamiltonian is 
the three-band model given by Liu {\it et al.} \cite{Liu2013threeband}, that describes, up to the next-nearest-neighbor order, the effective tight-binding model between the metal atoms M=\{Mo,W\}.

To calculate the equation of motion (\ref{mov_eq}), we again need the commutators $[\hat{\rho}_{0i}, \hat H^{3\mathrm{b}}_0]$, $[\hat{\rho}_{0i}, \hat H_I]$, and
$[\hat{\rho}_{0i}, \hat H_{ee}] $, $i=1,2$. The first two commutators give analogous results to those of the previous section, and the last one is given in equation (\ref{useful}). We are interested only in the term equivalent to the Bethe-Salpeter equation (encoded in the Rabi frequency renormalization term):
\bea
{\cal B}^{3\mathrm{b}}_{\mathbf{k}i}(\omega)=\frac{1}{S}\sum_\mathbf{q} V(|\mathbf{k}-\mathbf{q}|) F^{3\mathrm{b}}_{0i}(\mathbf{k},\mathbf{q}) p_{0i}(\mathbf{q},\omega),
\eea
which is obtained from equation (\ref{useful}) from the terms with $\lambda_1=0$, $\lambda_2=0,i$ and $\lambda_3=i$. For the model we are considering, the four-body
spinor is given by:
\be
 F^{3\mathrm{b}}_{0i}(\mathbf{k},\mathbf{q})=\phi^\dagger_0(\mathbf{k}+\mathbf{q}) \phi_0(\mathbf{k})\phi^\dagger_i(\mathbf{k}) \phi_i(\mathbf{k}+\mathbf{q}),
\ee
and is obtained numerically from the diagonalization of the $3\times3$ matrix defining  the three-band model, with $\phi_i$ the wave function of the $i$-band.

In our calculation, the exchange self-energy is included as an energy shift from the renormalization of the band gap [see equation (\ref{gap_analytic})]. We also ignore spin-orbit effects in the calculations at the $\Gamma$-point, as these are very small. The equation that
we need to solve for obtaining the transition probability $p_{0i}(k,\theta,\omega)$ reads:
\bea
\left[\omega-\omega_i(k,\theta)\right]p_{0i}(k,\theta,\omega)=-\Delta f_id^\Gamma_i(k,\theta){\cal E}_0+ \nonumber\\
\Delta f_i \frac{1}{S}\sum_{q,\theta^\prime} V(k,q,\theta-\theta^\prime)
F_i(k,q,\theta,\theta^\prime)p_{0i}(q,\theta^\prime,\omega),
\label{gamma_eq1} 
\eea	
with $\omega_i(k,\theta)=E_{ik,\theta}-E_{0k,\theta}$,  $E_{ik,\theta}$, and $E_{0k,\theta}$ the eigenvalues of the three-band Hamiltonian $H^{3\mathrm{b}}_0$, $\Delta f_i=f_{ik,\theta}-f_{0k,\theta}$ is the difference in occupation numbers between bands $i$ and $0$ (given in terms of 
the Fermi-Dirac function), and the dipole element $d^\Gamma_i(k,\theta)=\langle \lambda \mathbf{k}| \hat x| \lambda^\prime \mathbf{k} \rangle $, with the expectation value calculated using equation (\ref{dip_gamma}).

\begin{figure}
	\centering
	\includegraphics[scale=0.6]{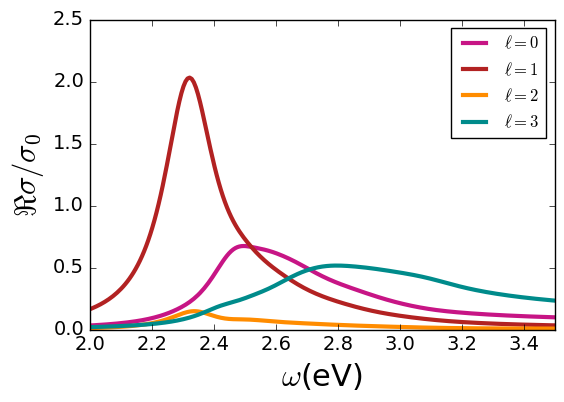}
	\caption{(Color on-line) Angular decomposition from each contribution to the optical conductivity for the exciton at the $\Gamma$-point for the $01$ exciton 
		(see section \ref{gamma} for the definition). In the calculation we have used the example of MoS$_2$.  This exciton is composed of Mo $d$-states. Each angular decomposition has also
		angular components $\ell\pm6$ and $\ell\pm12$, a consequence of the optical band structure (\ref{gamma_obs}).
	} \label{ang_dec}
\end{figure}

Since we only need the band-structure near the $\Gamma-$point we approximate the optical band structure
near this point by the Fourier series:
\be
\omega_i(k,\theta)\approx h^i_0(k)+h^i_6(k)\cos(6\theta)+h^i_{12}(k)\cos(12\theta), \label{gamma_obs}
\ee
with $h^i_\ell(k)$, $\ell=0,6,12$ are polynomials of degree six
(this expression is valid up to momentum values of the order of
$2\pi/(3a_0)$, where $a_0$ is the lattice parameter). Expression (\ref{gamma_obs}) describes
accurately  the optical band structure near the $\Gamma$-point. For $k\rightarrow\infty$ the optical band approximation (\ref{gamma_obs}) diverges. Although the contributions for $k\rightarrow\infty$ becomes negligible to the excitons' wavefunction, the approximation (\ref{gamma_obs}) makes the numerical convergence faster. Using the angular decomposition
(\ref{ang_dec}), $p_{i0} (k,\theta,\omega)={\cal E}_0\sum_\ell c_\ell^i(k) e^{\im\ell\theta} $ (note that we have omitted the dependence of $c_\ell^i(k)$ in $\omega$), we can write the Bethe-Salpeter equation (\ref{gamma_eq1}) as (where we have made
$\Delta f_i=-1$ since are interested in a neutral system):
\bea
 \omega c^i_\ell(k)-\sum_{\ell^\prime}\tilde{\omega}^i_{\ell^\prime}(k) c^i_{\ell-\ell^\prime}(k)=d^\Gamma_{i,\ell}(k)-
\sum_{\ell^\prime}\int q \frac{dq}{2\pi} \int \frac{d\theta}{2\pi}  \nonumber\\
\int \frac{d\theta^\prime}{2\pi} e^{-\im\ell \theta}e^{\im\ell^\prime \theta^\prime }   V(k,q,\theta-\theta^\prime)
F^{3\mathrm{b}}_{0i}(k,q,\theta,\theta^\prime)c_{\ell^\prime}^i(k), \label{gamma_1}
\eea
and $\tilde{\omega}^i_\ell(k)= \int_0^{2\pi} \frac{d\theta}{2\pi} \, e^{\im\ell \theta} \omega_i(k,\theta) $.

The previous equation couples coefficients $c_\ell^i$ with different angular momentum numbers $\ell$ through two terms: the
kinetic term $\sum_{\ell^\prime}\tilde{\omega}^i_{\ell^\prime}(k) c^i_{\ell-\ell^\prime}(k)$ and
the electron-electron interaction term (Rabi frequency renormalization term). The kinetic term couples only coefficients having
$\ell^\prime=0,\pm6,\pm12$, a consequence of the sixfold symmetry of equation (\ref{gamma_obs}). The potential term also couples coefficients $c^i_\ell(k)$ with
different angular momentum values, but this term gives a negligible contribution to the optical response when
$\ell\ne\ell^\prime$. This is a consequence of the fast vanishing of the potential $V(k,q,\theta-\theta^\prime)$ whenever $\theta-\theta^\prime \ne 0$ and $\theta-\theta^\prime $ is varied. Therefore, we can replace the four-body spinor function $F^{3b}_{0i}(k,q,\theta,\theta^\prime)$ by its
average angular value as follows:
\be
\tilde{F}^{3b}_{0i}(k,q,\theta^\prime)=\int_0^{2\pi} \frac{d\theta}{2\pi} F^{3b}_{0i}\left(k,q,\theta-\frac{\theta^\prime}{2},\theta+\frac{\theta^\prime}{2}\right). \label{pot_aprox2} 
\ee
and the effective potential is given in this approximation by:
\bea
\tilde{V}^i_{\ell}(k,q)=  \int_0^{2\pi} \frac{d\theta^\prime}{2\pi}
e^{\im\ell \theta^\prime }   V(k,q,\theta^\prime)
\tilde{F}^{3b}_{0i}(k,q,\theta^\prime). \label{pot_aprox}
\eea

The approximation (\ref{pot_aprox}) and (\ref{pot_aprox2}) keeps the hermiticity of  equation (\ref{gamma_1}). Finally, we replace the potential term in equation (\ref{gamma_1}) by (\ref{pot_aprox}), where we arrive at the coupled set of integral equations:
\bea
\omega c^i_\ell(k)-\sum_{\ell^\prime=0,\pm6,\pm12} \tilde{\omega}^i_{\ell^\prime}(k)& c^i_{\ell-\ell^\prime}(k)= d^\Gamma_{i,\ell}(k)-\nonumber\\
&-\int_0^\infty q dq \tilde{V}^i_\ell(k,q) c^i_\ell(q). \label{gamma_final}
\eea

To solve (\ref{gamma_final}), the summation in $\ell^\prime$ gives five additional terms [$\ell^\prime=0,\pm 6,\pm 12$; see equation (\ref{gamma_obs})] that are coupled together. 
This generates a hierarchy of equations for the coefficients $c^i_\ell(k)$.
Therefore,
the solution of equation (\ref{gamma_final}) has to be truncated 
at some $\ell$ value. In this procedure we have assumed that the contributions above 
$c^i_{18}(k)$ are vanishing small. This is confirmed by figure \ref{ang_dec}, which shows that
for $§\ell=5$ the contribution is already small (note that for this curve we have all the coefficients $c^i_{\ell}(k)$, with  $\ell=-7,-1,11,17$ entering the calculation of the conductivity, see below).
In terms of the coefficients $c^i_{\ell}(k)$ the conductivity is computed as follows.
The expectation value of the polarization operator can be calculated as we did in section \ref{formalism}, and results in:
\bea
P(\omega)&=&-2S\sum_{i=0,1}
 \int_0^\infty 
\int_0^{2\pi} k\frac{dk}{2\pi} \frac{d\theta}{2\pi}  \left[d^\Gamma_{i}(k,\theta)\right]^*\times\nonumber\\
&\times& p_{i0}(k,\theta,\omega),
\eea
where we account for the spin degeneracy introducing a factor of two. The conductivity can be obtained from equation (\ref{sigma_P_relation}), and we can separate the contribution for each band $i$. Performing the
angular integral in the equation for $P(\omega)$ we obtain:
\be
\frac{\sigma_i(\omega)}{\sigma_0}
=-8\im\omega \sum_{\ell=-\infty}^\infty \int_0^\infty k\frac{dk}{2\pi} \left[d^\Gamma_{i,\ell}(k)\right]^* c^i_\ell(k).
\ee
Once the coefficients $c^i_\ell(k)$ are determined from the solution of  (\ref{gamma_final}) the conductivity 
follows from the previous equation.

\begin{figure}
\centering
   \includegraphics[scale=0.6]{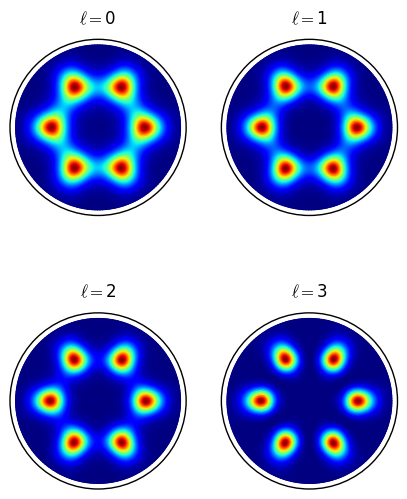}
   \caption{(Color on-line) Exciton wave function in momentum space at the $\Gamma$ point, for excitons composed of bands $0$ and $1$.
Note that all excitons have nodes along the $\Gamma-\mathbf{M}$ direction (vertical).  The differences between the different wave functions are subtle.   
} \label{gamma1}
\end{figure}

\begin{figure}
\centering
   \includegraphics[scale=0.6]{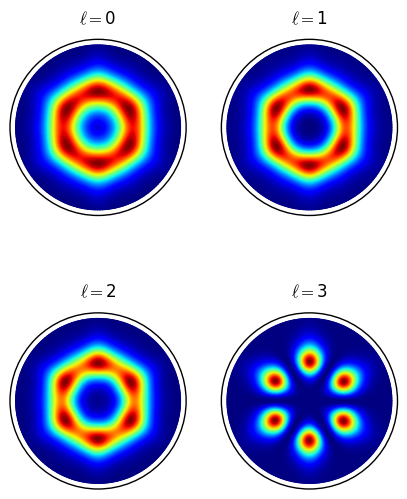}
   \caption{(Color on-line) Exciton wave function  in momentum space at the $\Gamma$ point, for excitons composed of bands $0$ and $2$.
Note that the  all exciton have nodes along the $\Gamma-\mathbf{K}$ direction (horizontal).   
} \label{gamma2}
\end{figure}

The solution of (\ref{gamma_final})  also give us the excitonic wave functions in momentum space.
The results for the first excitonic energy, for each angular momentum mode, is shown in figure \ref{gamma1}  for the exciton
composed from an electron in band $1$, and in figure \ref{gamma2} for an electron in band $2$. From a careful inspection of figures \ref{gamma1} 
and \ref{gamma2}, we can see that the nodes of  exciton with band index $i=1, \ell=0$ lies along the $\Gamma-\mathbf{K}$ direction, while the nodes of  exciton with band index $i=2, \ell=0$ lies along the $\Gamma$-M point.

\section{Results \label{results}}

In this section we perform a thorough analysis of the absorption spectrum
of four TMDC's. 
For computing the absorbance, the optical conductivity is needed.
Taking the example of MoS$_2$, the 
decomposition of the 
real part of the optical conductivity, coming from different angular momentum contributions of the exciton at  $\Gamma$-point, associated with the transition $0\rightarrow1$, is shown in figure \ref{ang_dec}; remember that each contribution is composed of $\ell=0$, $\ell\pm6$, and $\ell\pm12$ angular momentum components.

It  is important to introduce here a note on notation: the peak at lowest energy is denoted by A$=1s$ and the next Rydberg energy level in the A-series is denoted by A'$=2s$; this corresponds in a given valley and to a given spin projection.
In the same valley, and for the other spin projection, the peaks belong to the B-series, with the lowest energy is denoted by B=$1s$ and the next one by B'$=2s$. For MoX$_2$ TDMC's the energy order is A, B, A', and B', whereas for WX$_2$ TMDC's the energy order is A, A', B, and B'. This agrees with the notation introduced in figure \ref{exciton_exchange_bse}.

\begin{figure*}
	\centering
	\includegraphics[scale=0.6]{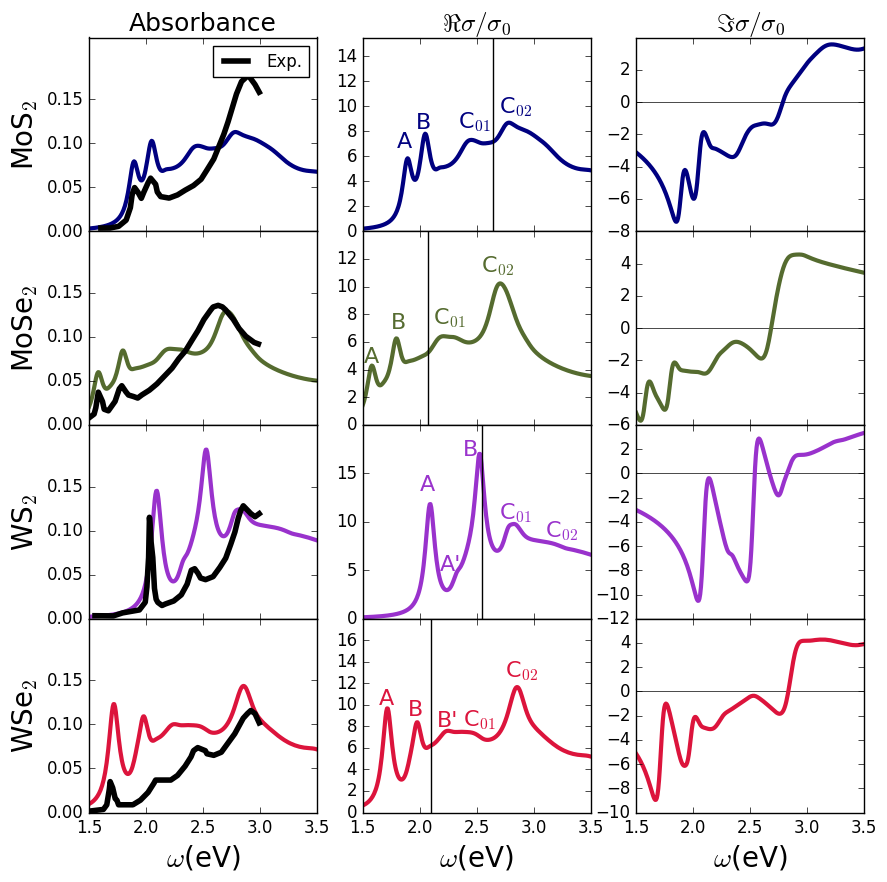}
	\caption{(Color on-line) Absorbance and optical conductivity of four TMDC's computed from formula (\ref{eq_absorb}), assuming the materials are neutral and in vacuum; there is no fitting of the data.  The real part of the conductivity has the peaks labeled by the
		corresponding excitonic series, A (1$s$=A and 2$s$=A'), B (1$s$=B and 2$s$=B'), and C (all contributions from figure \ref{ang_dec} for the transitions 0$\rightarrow1$ and 0$\rightarrow2$). The former two are due to transitions at the \textbf{K}-point and the latter 
		to transition at the $\Gamma$-point. Experimental data for the absorbance (solid black curves) is taken from reference \cite{li2014measurement}. The model parameters are given in table \ref{parameters_table} for the excitons at the \textbf{K}-point and in reference \cite{Liu2013threeband} for the $\Gamma-$point; for all but WS$_2$ the parameter 
		$r_0$ has been replaced by $r_0\epsilon$, with $\epsilon$ the effective dielectric constant for a fused silica substrate; a similar procedure was used in reference
		\cite{WS2exp2}.
		See section \ref{results}  for a discussion of the similarities and differences between the data and the computed spectra.  
		The vertical black line in the central panels define the energy value 
		above which we enter the continuum of the A-series. Note that for WX$_2$ the first peak of the B series in inside the continuum of the A series (this accounts for the disappearance of the B peak in the experimental data; see figure \ref{absorbance}).
		Excitons in the A and B series corresponding to $\ell=-2$ have vanishing contribution to the optical properties at exciton energy, but are included in this calculation
		(note that $\ell=2$ is a dark exciton). The absorbance has been computed taking the substrate into account using the dielectric permitivity of fused silica at optical frequencies ($\epsilon=2.13$).
	} \label{general}
\end{figure*}

The absorbance, and the real and the imaginary parts of the optical conductivity, for four TMDC's considered in this work, are shown in figure \ref{general}, with the parameters of table \ref{parameters_table}. That is, in this figure we do not try to fit the data but simply use the parameters
characterizing the potential and the band-structure of the TMDC's given in other papers.
In figure \ref{absorbance}, on the contrary, we fit the A peak position changing $r_0$ and we
also add a chemical potential, since,  as noted in Refs. \cite{mak2013tightly,chernikov2015} 
all  TMDC's samples have a certain and undetermined amount of negative doping.
We note in passing that at the time of writing different experiments report
distinct  percentages for  absorption of radiation for two, supposedly identical, TMDC's. Table \ref{tab:absexp} gives, from four different references,
the measured values of the absorbance of MoS$_2$ samples; as it can be seen the values fluctuate among different experiments. Also, our model predicts, at low temperatures,
larger absorption peaks than those measured at room temperature. This result makes sense, but when we increase the temperature we never obtain values as small as those reported in the experiments for MoS$_2$. 
It is now known \cite{Mak2017} that excitonic spectrum of TDMC's samples in SiO$_2$ are strongly influence by 
the disorder of the substrate. In this reference it is shown that encapsulated samples in h-BN
have much narrower excitonic peaks. Therefore our results should agree with absorbance measurements in these encapsulated samples (measurements yet to be made).

\begin{table}
\begin{tabular}{cccccc}
 excitonic peak &Model &   \cite{mak2010atomically} &   \cite{li2014measurement} &  \cite{Morozov2015} &  \cite{mak2013tightly}\\
 \hline
A& 14 & 10.8 &7.4 & 3.8 &7.5 \\
B& 15 & 10.5& 8.6& 5.0 & 8.0\\
\hline
\end{tabular}
\caption{Absorbance (in percentage) of A and B peaks for MoS$_2$. The table gives a comparison between our theoretical model and the  experimental results for  samples deposited on silica (\cite{mak2010atomically},\cite{li2014measurement},\cite{Morozov2015}),  and a FET device \cite{mak2013tightly}, where the MoS$_2$ is deposited on top of silicon and under a voltage gate of $-10$ V. 
The ``Model" refers to the theoretical approach developed in this paper and
we have considered  MoS$_2$ on top of silica ($\epsilon_{\rm{silica}}=1.46$). This value
of $\epsilon_{\rm{silica}}$ translates into an $\epsilon_m=(1+2.13)/2$, which is the 
value we use in our equations.
Remember that in our model  both the peak intensity and the peak width
are dependent on the choice of the relaxation rate $\gamma_K$. We can artificially reduce the height of the peak at the expenses of 
increasing its width. It is worth noting the variation of the experimental values
for the absorbance among themselves.
\label{tab:absexp}}
\end{table}

Next,  we  analyze each aspect of the optical spectrum
of each TMDC
 and compare our results with the experimental measurements available to date
 in a large frequency window. The parameters used in our calculations are: (i) at the $\mathbf{K}$-point we used the values in table \ref{parameters_table} and a broadening $\gamma_K=50$ meV;
 (ii) at the $\Gamma$-point we used the GGA parameters of the three-band model
 given by Liu {\it et al.} \cite{Liu2013threeband}, the same Keldysh parameters of table \ref{parameters_table}, and a broadening $\gamma_\Gamma=100$ meV.
 Note that exception made to the broadening parameters, all the other values
 were taken of the literature and no attempt was made to choose them in order to fit the data, with exception to the case reported in figure \ref{absorbance}.

\begin{itemize}

\item \textbf{MoS}$_2$
The two first peaks in the absorbance, A and B, correspond 
to the A($1s$) and B($1s$) excitons. The different position of the two peaks is a consequence of SOC splitting of the bands. 
The last two peaks in the absorbance spectrum, 
$C_{01}$ ans $C_{02}$ (having about the same intensity --see the conductivity curve),
correspond to the sum of different angular momenta contributions from the $\Gamma$-excitons (actually excitonic resonances). The third
($C_{01}$) peak is associated with the transition from the top of the valence band to both the 1 and 1' conduction bands at the $\Gamma$-point
(conduction bands number 5 and 6 in figure \ref{MoS2_band_structure}); we have considered these two degenerated since SOC is small in this case. Finally,  the fourth peak ($C_{02}$) comes from transitions connecting the top of the valence band and the  2 and 2' conduction bands (also taken degenerated; conduction bands 7 and 8 in figure \ref{MoS2_band_structure}). 

The real part of the conductivity follows closely the absorbance spectra, as expected. Usually, the imaginary part of the conductivity from
a single excitonic contribution is negative for $\hbar\omega< E_b$ and positive for $\hbar\omega>E_b$,  where $E_b$ is the binding energy, a result that can be obtained by  inspection of Elliot's formula for the optical conductivity (\ref{elliot_cond}). 

Let us now discuss the differences between  experimental data and our model. We note that the rigid shift to the left  performed  
by Wu {\it et al.}
 \cite{Wu2015},
and Steinhoff  {\it et al.} \cite{steinhoff2014influence} is not necessary in our case. The difference in  intensity of  A and B peaks is probably a consequence of the phonons that exist at finite temperature. This effect was not considered in this work but was shown
 to be important for the peaks's broadening \cite{qiu2013optical,molina2016temperature}. 

Lastly, we discuss the excitonic effects at the $\Gamma$-point. The optical measurements identify only one peak, which seems to 
  correspond to the $C_{02}$ exciton. The
work of Qiu \textit{et al.} \cite{qiu2013optical}  obtain a rich structure of peaks in this region that was washed out when they include quasi-particle lifetimes from phonon terms.

\item \textbf{MoSe}$_2$

The aborbance spectrum of this TMDC share many similarities with  MoS$_2$: two peaks from the $\mathbf{K}$-point split (A and B) by the SOC and two wider peaks from the $\Gamma$-point are also present.  The exciton at the $\Gamma$-point contributes with two peaks at $\sim2.4$ eV and $\sim2.7$ eV. The experimental data shows a single peak at $2.6$ eV. This discrepancy  comes possibly   from the phonons already discussed for the MoS$_2$. Overall there is a good agreement between the data and the calculated curves, both in position of the peaks and in intensity.

The imaginary part of the conductivity is only positive for frequencies $\hbar\omega>2.6$ eV, meaning that exciton-polaritons can only be excited for energies in the visible.

\item \textbf{WS}$_2$

For this material we note the very good agreement of the position 
and magnitude of the calculated A peak in comparison with those in the experimental data. We also see  that the second experimental peak coincides with a small computed peak from the 2s  state (A') associated with the series of first exciton A-peak (see figure \ref{exciton_exchange_bse}). 
There is at least three reports \cite{HanbickiBpeak,ZhuBpeak,WS2Bpeak} of measurement of the A' peak in WS$_2$ in  the temperature range of 4-300 K.
Unfortunately, in the literature the A' peak has been dubbed B, using an analogy
with the MoX$_2$ case. 
However, looking at the central panel of figure \ref{general} we clearly see that the A'$=2s$ peak appears at lower energy than the B$=1s$.
Note that from our analysis we can separate each spin/valley contribution.
Also note that the  A' peak has a similar absorbance to the experimental one 
(identified in the experimental literature on WX$_2$ TMDC's as B, because it is the second to appear in the energy scale). Studying the dependence of light absorption of different peaks on an external magnetic field, for breaking spin degeneracy, together with the use of strong circular polarized light to populate the two valleys differently\cite{Anshuman2016}, is a possible way of clarifying the microscopic origin of the different peaks.

The third theoretical (B) peak (which is the SOC counterpart of the first peak) is absent in the experimental data. Note that from figure \ref{exciton_exchange_bse}, all but one (1$s$=B) contributions  from the B family of peaks are excitonic resonances (above the interacting band gap). The proximity of the B-peak to the continuum may provide a scattering channel to transfers
spectral weight from this peak to the resonances in the continuum. An additional and possible mechanism is based on extrinsic doping of these materials as shown in figure \ref{absorbance}. It has been shown that doping has a strong effect in attenuating the 
excitonic peaks \cite{gao2016dynamical}, specially the B-peak in MoS$_2$. There is no reason to believe that
the same mechanism would not work in WX$_2$ TMDC's.  Indeed, from figure \ref{fig:spin_bands} we expect a strong attenuation of the high-energy excitonic peak whereas the low energy one should survive. This should happen since the doping with electrons tends to block first the higher energy transition whereas maintaining the low energy one. In figure \ref{absorbance} we see a
	comparison of the absorption spectrum of WS$_2$ with the data taking into account the effect of doping; the agreement is excellent. The suppression of the B-peak is evident from our results, thus confirming  doping by electrons as a possible mechanism for suppressing the B excitonic state.

The first excitonic resonance ($C_{01}$) at the $\Gamma$-point is in very good agreement with the experimental one, while the second excitonic  ($C_{02}$)  resonance at the $\Gamma$-point is at an energy range above the measured one (although its intensity is rather small). Therefore nothing can be said about the possible agreement with the experimental data, since this does not cover that spectral region. Lastly the imaginary part of the optical conductivity becomes positive above the energy $\sim2.6$ eV nd therefore the system  can support exciton-polaritons in that spectral region. 

\item \textbf{WSe}$_2$

We end our analysis with a comparison between the  calculated absorbance curve
 and the one measured for WSe$_2$. For this material the disagreement between the calculated curves and the experimental data is the largest of the four TMDC's studied in this work.
 Indeed, the data seems stretched relatively to the calculated curves.
The first peak in the WSe$_2$ absorbance spectrum is  in very good agreement with the experimental data, with a difference in position less than $0.1$ eV. 

 As in
the case of  WS$_2$, we see that  the B-family peak is present in the data as a small shoulder. 
The third and fourth experimental peaks, when compared with our theoretical model, come from resonances at the $\Gamma$-point. The theoretical calculations show a 
 red shift of about $0.2$ eV for these two peaks, indicating that higher order exchange corrections,  which reshape the band structure around the $\Gamma$-point, might be important.

The imaginary part of the optical conductivity is positive above $\sim2.8$ eV, thus allowing for excitons-polaritons.
\end{itemize}


Next we present an analysis of the effects associated with changing the Fermi energy and the Keldysh potential parameter $r_0$. Given a Fermi energy the parameter $r_0$ can be adjusted to fit the A peak. Results of this procedure are shown in figure
(\ref{absorbance}) for  WS$_2$. We can see an excellent  agreement between our results and the experimental curve. This highlights the importance of a finite Fermi energy in describing the experimental data. As noted before a finite Fermi energy comes  
from the spontaneous negative doping observed
in TMDC's samples. The better agreement with the data shown in figure (\ref{absorbance}) relatively to  the results of figure (\ref{general}) shows
the non-negligible effect of the doping in the  optical properties. On the other hand, the parameter $r_0$ should also be a function of the electronic density. At the moment of writing this dependence is 
unknown.

One aspect that our calculation does not take into account in an exact way is the self-consistent solution of the exchange energy. Since this calculation is outside the scope of this work,  we can mimic it using a different value of $\Lambda_2$ [see Eq. (\ref{lambda2})]. This leads to a narrow A peak and a broaden B peak in WS$_2$, as 
seen in the experimental data. In this regime, the B peak is no longer an exciton but rather an excitonic resonance.  The mechanism
leading to the broaden of the B peak can be explained by the self-consistent solution of the exchange energy. For a given carrier density, the iterative calculation of the exchange energy reduces its value and therefore the importance of the doping increases for the lowest conduction band. In WS$_2$ the effect is much stronger in the lowest band than in the next conduction band due to the large spin-orbit splitting. This mechanism due to exchange increases the splitting  on the two conduction bands.

\begin{figure}
	\centering
	\includegraphics[scale=0.6]{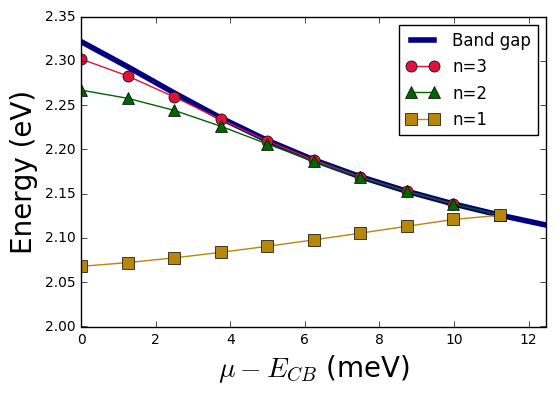}
	\caption
	{ Dependence of the band gap and of the three first exciton s-states for WSe$_2$ on the Fermi energy. $E_{CB}$ is the lowest
		conductance band energy. The parameters of the mass, Fermi velocity and SOC are from table \ref{bse_table}. We set $r_0=40.92\,\,$ \AA,
		slightly larger than that given in reference \cite{Berkelbach2013}, $37.89\,\,$\AA.
		The effective dielectric constant, including the effect of the substrate (SiO$_2$), is $\varepsilon=2.45$.
		We used a temperature of 77 K.
	} \label{ef_dependence}
\end{figure}

Another aspect of the doping is its influence on the decreasing of the band gap.
We show in figure (\ref{ef_dependence}) the dependence of the band gap and the exciton energies on the Fermi energy. We can see
that increasing Fermi energy makes the binding energy (difference between the 
thick blue curve and all the others) smaller. The energy of the first excited state (squares) increases with the doping while the energy 
the second (triangles) and third (circles)  have the opposite behavior. We also see that it exists a critical doping that makes the exciton states collapsing into resonances when they merge with the band gap. For the energy of first and third excited states we see the same qualitative behavior as measured in Ref. \cite{chernikov2015}.


This concludes the analysis of our theoretical results when compared with the 
experimental data. Globally, the agreement is good, but some points need further research.
Measurements performed at low temperatures in encapsulated TMDC's
using hexagonal boron-nitride  should reveal the fine        structure of the excitonic spectrum
predicted by our model. 
\section{Discussion and conclusions\label{conclusions}}

In summary, we performed a study of excitons in TMDC's  monolayers 
 including in the same foot both excitons at the $\mathbf{K}$- and $\Gamma$-points. The excitons at the
$\mathbf{K}$-point were calculated with a gapped Dirac equation including electron-electron interactions and SOC. The excitonic resonances at the $\Gamma-$point
were calculated with the tight-binding three-band model expanded around that point in the Brillouin zone.
We  compared our theoretical results with the experimental data available 
from reference \cite{li2014measurement}. We clarified the  microscopic origin of each
observed excitonic peak and discussed the reasons for some disagreement between our theoretical model and the experimental data. Note that the 
measurements where made at room temperature. Therefore, the effect of 
a self-energy, which will be energy dependent, due to electron-phonon interactions might play an important role in modeling the absorbance 
spectrum at room temperature. We note here that our equation of motion method also allows for treating electron-phonon interactions at the expenses of a more lengthly calculation. 

\begin{figure}
\centering
   \includegraphics[scale=0.6]{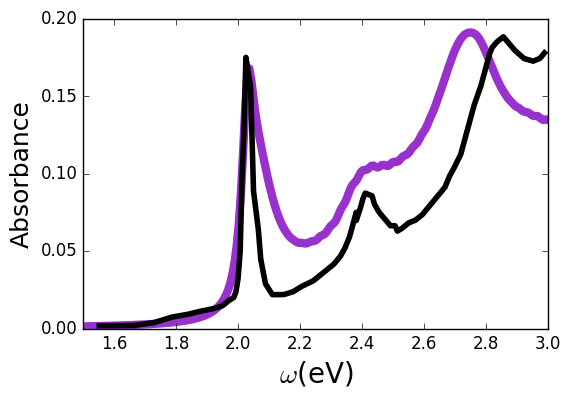}
  \caption{ Fitting of the  experimental optical absorption for WS$_2$. The parameters are those of table \ref{parameters_table} unless otherwise said. The new parameters are $r_{0,A}=55.7\AA$, $E_F=5$ meV,   $\gamma_A=\gamma_B=26$ meV, and $\gamma_\Gamma=0.1$ eV. Temperature is 300 K and we also changed the value of $\Lambda_2\rightarrow\Lambda_2+0.12$.
  	The figure is discussed in detail in the main text.
  } \label{absorbance}
\end{figure}

Also, as noted by Mak {\it et al.} \cite{mak2013tightly}: ``Spontaneous negative doping, presumably from defects within the MoS$_2$ layer
and/or substrate interactions, has been commonly reported in mechanically exfoliated samples". 
This seems be the reason \cite{gao2016dynamical} why the B-series is not visible in WX$_2$ when the material is electron-doped
(see figure \ref{absorbance}). 
To conclude, given the uncertainties in the experimental  data 
reported in table \ref{tab:absexp}
we consider the agreement between our calculation and the data to be quite good.

We have also studied the variation of the $1s=$A peak with the dielectric function of the capping medium (results not shown).
We found that the A-peak position varies little with $\epsilon_m$. This happens because the exchange energy correction compensates the binding energy  coming from the BSE.

Although we have considered in this work the response to linearly polarized electromagnetic radiation, it is simple to generalized the calculation to circularly 
polarized one. This would allow us to discuss the additional appearance of 
more selection rules associated with spin.

Finally, we note that our method can be easily generalized to the calculation of excitonic effects in other 2D materials, such as phosphorene, a line of research 
we will pursue in a forthcoming paper. This case will be particularly interesting due to the strong anisotropy of the material.

A future working direction is the inclusion of  density and non-linear terms coming from the commutator of the density matrix with the electro-electron interaction. In particular the latter term will be important for the discussion
of excitonic-assisted nonlinear optics in TMDC's
(see also \cite{Pedersen2015}). Note that contrary to 3D semi-conductors the optical spectrum is dramatically affected by 
excitonic excitation even for frequency ranges well above the non-interacting gap. Thus, a description of the nonlinear optics
properties of TMDC's using a single electron picture is doomed to fail.

\section*{Acknowledgments}
A. J. Chaves acknowledges a scholarship from the Brazilian
agency  CNPq (Conselho Nacional de Desenvolvimento
Cient\'ifico e Tecnol\'ogico).
 N.M.R. Peres and R.M. Ribeiro acknowledge support from the European Commission through
the project ``Graphene-Driven Revolutions in ICT and Beyond"
(Ref. No. 696656), project PTDC/FIS-NAN/3668/2014, and the Portuguese Foundation for Science and Technology 
(FCT) in the framework of the Strategic Financing UID/FIS/04650/2013. 
T. Frederico thanks for the support of the Brazilian Agencies CNPq and FAPESP (Funda\c c\~ao de Amparo \`a Pesquisa do Estado de S\~ao Paulo).
The authors acknowledge Fanyao Qu and Alan MacDonald for useful discussions about the numerical solution of the homogeneous Bethe-Salpeter equation, and Ermin Malic for sharing with us useful supplementary notes.

\begin{appendix} 
\section{Calculation of  the commutators} \label{commutators} \label{auxiliar_functions}

The equation of motion for density matrix $\hat{\rho}_{\lambda\lambda^\prime}^{s\tau}$ will be
determined in this appendix. We need do calculate the following commutators: $[\hat H_0,\hat{\rho}^{s\tau}_{\lambda \lambda^\prime}((\mathbf{k},t)]$, $[\hat H_\mathrm{ee},\hat{\rho}^{s\tau}_{\lambda \lambda^\prime}(\mathbf{k},t)]$ and $[\hat H_I,\hat{\rho}^{s\tau}_{\lambda \lambda^\prime}(\mathbf{k},t)]$. 
For the first of these we have:
\be
[\hat H_0,\hat{\rho}^{s\tau}_{\lambda,\lambda^\prime}(\mathbf{k},t)]=(\lambda-\lambda^\prime)E^{s\tau}_k \hat{\rho}^{s\tau}_{\lambda,\lambda^\prime}(\mathbf{k},t)\,. \label{H0_term}
\ee
For the external field, we only consider the interband terms of the Hamiltonian $\hat H_I$ as we are discussing a neutral system. In this condition $\hat H_I$ reads:
\be
\hat H_I=-\im e {\cal E}(t) \sum_{s\tau\lambda_1\mathbf{k}} \frac{v^{s\tau}_{\lambda_1}(\mathbf{k})}{2\lambda_1  E^{s\tau}_k }  \hat \rho^{s\tau}_{\lambda_1-\lambda_1} (\mathbf{k},t),
\ee
and therefore:
\bea
[\hat H_I,\hat{\rho}^{s\tau}_{\lambda,\lambda^\prime}(\mathbf{k},t)]=\im e {\cal E}(t)& \Bigg[  \frac{v^{s\tau}_{-\lambda}(\mathbf{k})}{2\lambda E_k^{s\tau}} \hat{\rho}^{s\tau}_{-\lambda,\lambda^\prime}(\mathbf{k},t)+\nonumber\\
&+\frac{v_{\lambda^\prime}^{s\tau}(\mathbf{k})}{2\lambda^\prime E_k^{s\tau}} \hat\rho^{s\tau}_{\lambda,-\lambda^\prime}(\mathbf{k},t) \Bigg],
\eea	
with the corresponding expectation values:
\be
\left\langle[\hat H_I,\hat{\rho}^{s\tau}_{\lambda,\lambda}(\mathbf{k},t)]\right\rangle= \im e {\cal E}(t)  \frac{\Im \left[ \im v^{s\tau}_{\lambda}(\mathbf{k})p^{s\tau}_{\lambda}(\mathbf{k},t) \right]}{\lambda E_k^{s\tau}} , \label{n_HI_term}
\ee	
and
\bea
\left\langle[\hat H_I,\hat{\rho}^{s\tau}_{\lambda,-\lambda}(\mathbf{k},t)]\right\rangle=\im e {\cal E}(t) \frac{v^{s\tau}_{-\lambda}(\mathbf{k})}{2\lambda E_k^{s\tau}} \times\nonumber\\
\left[n^{s\tau}_{-\lambda}(\mathbf{k},t)- n^{s\tau}_{\lambda} (\mathbf{k},t)\right]. \label{p_HI_term}
\eea
Finally,  the commutator with the electron-electron interaction reads:
\bea
[\hat H_\mathrm{ee},\hat{\rho}^{s\tau}_{\lambda \lambda^\prime}(\mathbf{k},t)]=\frac{S}{2}\sum_{\mathbf{k}^{\prime}\mathbf{k^{\prime\prime}},\mathbf{q}\ne \mathbf{0}} \sum_{\lambda_1\lambda_2\lambda_3\lambda_4}\sum_{s_1\tau_1 s_2 \tau_2}\nonumber\\
\left[\phi^{s_1\tau_1}_{\lambda_2}(\mathbf{k}^{\prime\prime}+\mathbf{q})\right]^\dagger \phi^{s_1\tau_1}_{\lambda_3}(\mathbf{k}^{\prime\prime})
\left[\phi^{s_2\tau_2}_{\lambda_1}(\mathbf{k}^\prime-\mathbf{q})\right]^\dagger \phi^{s_2\tau_2}_{\lambda_4}(\mathbf{k}^\prime)
V(\mathbf{q})\Big[\nonumber\\
\hat a^\dagger_{\mathbf{k}^\prime-\mathbf{q}\lambda_1\sigma_1} \hat a^\dagger_{\mathbf{k^{\prime\prime}}+\mathbf{q}\lambda_2\sigma_2}\hat a_{\mathbf{k}^{\prime\prime}\lambda_3\sigma_2}a_{\mathbf{k}\lambda^\prime\sigma}\delta_{\lambda\lambda_4}\delta_{\sigma\sigma_1}\delta_{\mathbf{k},\mathbf{k}^\prime}+\nonumber\\
\hat a^\dagger_{\mathbf{k}^\prime-\mathbf{q}\lambda_1\sigma_1} \hat a^\dagger_{\mathbf{k}^{\prime\prime}+\mathbf{q}\lambda_2\sigma_2} \hat a_{\mathbf{k}\lambda^\prime\sigma}\hat a_{\mathbf{k}^\prime\lambda_4\sigma_1} \delta_{\mathbf{k},\mathbf{k}^{\prime\prime}} \delta_{\lambda \lambda_3} \delta_{\sigma_2\sigma}+\nonumber\\
\hat a^\dagger_{\mathbf{k}\lambda\sigma} \hat a^\dagger_{\mathbf{k}^\prime-\mathbf{q}\lambda_1\sigma_1} \hat a_{\mathbf{k}^{\prime\prime}\lambda_3\sigma_2} \hat a_{\mathbf{k}^\prime\lambda_4\sigma_1}\delta_{\mathbf{k}^{\prime\prime}+\mathbf{q},\mathbf{k}}\delta_{\lambda_2\lambda^\prime}\delta_{\sigma_2\sigma}+\nonumber\\
-\hat a^\dagger_{\mathbf{k}\lambda\sigma} \hat a^\dagger_{\mathbf{k^{\prime\prime}+q}\lambda_2\sigma_2} \hat a_{\mathbf{k}^{\prime\prime}\lambda_3\sigma_2}
\hat a_{\mathbf{k}^\prime \lambda_4 \sigma_1} \delta_{\mathbf{k}^\prime-\mathbf{q},\mathbf{k}}\delta_{\lambda_1\lambda^\prime}\delta_{\sigma_1\sigma}\Big]. \label{big_eq}
\eea
The  expectation value of four body operators is truncated at the RPA level:
\bea
\left\langle \hat a^\dagger_{\mathbf{k+q}\lambda_1\sigma_1} \hat a^\dagger_{\mathbf{k^\prime-q}\lambda_2\sigma_2}\hat a_{\mathbf{k}\lambda_3\sigma_2}\hat a_{\mathbf{k}^\prime\lambda^\prime\sigma}\right\rangle\approx \nonumber\\
\left\langle \hat a^\dagger_{\mathbf{k^\prime-q}\lambda_2\sigma_2}\hat a_{\mathbf{k}\lambda_3\sigma_2}\right\rangle\left\langle \hat a^\dagger_{\mathbf{k+q}\lambda_1\sigma_1}\hat a_{\mathbf{k}^\prime\lambda^\prime\sigma}\right\rangle,
\eea
and within this approximation, the expectation value of equation (\ref{big_eq}) is given by
($S=L^2$, the area of the system):
\bea
\left\langle[\hat H_\mathrm{ee},\hat \rho^{s\tau}_{\lambda\lambda^\prime}(\mathbf{k})] \right\rangle=S\sum_{\mathbf{q}} V(\mathbf{q})\sum_{\lambda_1\lambda_3}
\left\langle \hat \rho^{s\tau}_{\lambda_1\lambda_3}(\mathbf{k}-\mathbf{q})\right\rangle
 \times
\nonumber\\
\sum_{\lambda_2} \left[ F^{s\tau}_{\lambda^\prime \lambda_3 \lambda_1\lambda_2}(\mathbf{k},\mathbf{k}-\mathbf{q}) \left\langle  \hat \rho^{s\tau}_{\lambda \lambda_2}(\mathbf{p})\right\rangle 
-\right.\nonumber\\
\left.
F^{s\tau}_{\lambda_2\lambda_3\lambda_1\lambda}(\mathbf{k},\mathbf{k}-\mathbf{q})
\left\langle  \hat \rho^{s\tau}_{\lambda_2\lambda^\prime}(\mathbf{p})\right\rangle 
\right], \label{useful}
\eea
where we have used the property $V(\mathbf{q})=V(-\mathbf{q})$.  Equation (\ref{useful}) is valid for a system composed of \textit{any number
of bands}. For the particular case of two band systems, as in the case given by the Hamiltonian (\ref{full_H}), we split equation (\ref{useful}) into various terms. For $\lambda=\lambda^\prime$ we have:
\bea
\left\langle[\hat H_\mathrm{ee},\hat \rho^{s\tau}_{\lambda\lambda}(\mathbf{k})] \right\rangle=2iS \sum_\mathbf{q} V(\mathbf{k}-\mathbf{q}) \Bigg\{\nonumber\\
(n^{s\tau}_\lambda(\mathbf{q},t)-n^{s\tau}_{-\lambda}(\mathbf{q},t)\Im\left[F^{s\tau}_{\lambda\lambda\lambda-\lambda}(\mathbf{k},\mathbf{q})p^{s\tau}_{\lambda}(\mathbf{k},t) \right]+ \nonumber\\
+ \Im\Big[p_{\lambda}^{s\tau}(\mathbf{k},t) \Big(p_\lambda^{s\tau}(\mathbf{q},t)
 F^{s\tau}_{\lambda-\lambda\lambda-\lambda}(\mathbf{k},\mathbf{q})+\nonumber\\
+p_{-\lambda}^{s\tau}(\mathbf{q},t) F^{s\tau}_{\lambda\lambda-\lambda-\lambda}(\mathbf{k},\mathbf{q}) \Big) \Big]\Bigg\}, \label{n_Hee_term}
\eea
and for $\lambda^\prime=-\lambda$ we find:
\be
\left\langle[\hat H_\mathrm{ee},\hat \rho^{s\tau}_{\lambda-\lambda}(\mathbf{k})] \right\rangle=S \sum_{\mathbf{q}\lambda_1,i=1,..4}V(\mathbf{q})
X_i,\label{p_Hee_term}
\ee
\bea
X_1=p^{s\tau}_\lambda(\mathbf{k},t)n_{\lambda_1}^{s\tau}(\mathbf{k}-\mathbf{q},t)\times \nonumber \\
\Big[F^{s\tau}_{-\lambda\lambda_1\lambda_1-\lambda}(\mathbf{k},\mathbf{k}-\mathbf{q})-F^{s\tau}_{\lambda\lambda_1\lambda_1\lambda}(\mathbf{k},\mathbf{k}-\mathbf{q}) \Big]
\eea
\be
X_2=p_{\lambda_1}^{s\tau}(\mathbf{k}-\mathbf{q},t) F^{s\tau}_{-\lambda-\lambda_1\lambda_1\lambda}(\mathbf{k},\mathbf{k}-\mathbf{q})\Delta n^{s\tau}_\lambda(\mathbf{k},t)+
\ee
\bea
X_3=p^{s\tau}_\lambda(\mathbf{k},t)p^{s\tau}_{\lambda_1}(\mathbf{k}-\mathbf{q},t) \Big[F^{s\tau}_{-\lambda-\lambda_1\lambda_1-\lambda}
-\nonumber \\ -F^{s\tau}_{\lambda-\lambda_1\lambda_1\lambda}(\mathbf{k},\mathbf{k}-\mathbf{q}) \Big]
\eea
\be
X_4=n_{\lambda_1}^{s\tau}(\mathbf{k}-\mathbf{q},t)F_{-\lambda\lambda_1\lambda_1\lambda}^{s\tau}(\mathbf{k},\mathbf{k}-\mathbf{q})\Delta n_\lambda^{s\tau}(\mathbf{k},t).
\ee
Finally we need to add (\ref{H0_term}), (\ref{n_HI_term}), and (\ref{n_Hee_term}) to obtain the equation of motion for
$n^{s\tau}_\lambda$, and (\ref{H0_term}), (\ref{p_HI_term}), and (\ref{p_Hee_term}) to obtain the equation of motion
for $p^{s\tau}_\lambda$.

\section{Overlap of the four-body wavefunctions} \label{overlap_wavefunctions}
The four-body overlap functions are explicitly defined below for the 
massive Dirac Hamiltonian:

\bea
F^{s\tau}_{\lambda_1,\lambda_2,\lambda_3,\lambda_4}(\mathbf{k_1},\mathbf{k_2})=
\nonumber\\
={{\phi^{s\tau}_{\lambda_1}}^\dagger}(\mathbf{k}_1)
\phi^{s\tau}_{\lambda_2}(\mathbf{k_2})
{{\phi^{s\tau}_{\lambda_3}}^\dagger}(\mathbf{k_2})
\phi^{s\tau}_{\lambda_4}(\mathbf{k}_1)\,. \label{F_definition_ap}
\eea
For simplicity of writing, we omit in this appendix the superscript $s\tau$
in the $F$'s-functions and in the energy $E_k^{s\tau}$.
For the case $\lambda_1=\lambda_4$ and $\lambda_2=\lambda_3$
the overlap function reads:
\be
F_{\lambda_1,\lambda_2,\lambda_2,\lambda_1}(\mathbf{k_1},\mathbf{k_2})=\frac{1}{2}\left(1+\lambda_1\lambda_2 \frac{\mathbf{k_1}\cdot\mathbf{k_2}+m^2}{E_\mathbf{k_1}E_\mathbf{k_2}} \right), \label{F_pppp}
\ee
whereas when $\lambda_1=\lambda_4, \lambda_2=-\lambda_3$
we find:
\bea
F_{\lambda_1,\lambda_2,-\lambda_2,\lambda_1}(\mathbf{k_1},\mathbf{k_2})=\nonumber \\
\frac{\lambda_1}{2}\frac{ m\left[\mathbf{k_2}\cdot(\mathbf{k_2-k_1})\right]+i\lambda_2 E_\mathbf{k_2} ( \mathbf{k_1}\times \mathbf{k_2}) \cdot\mathbf{u}_z}{ k_2 E_{\mathbf{k_1}}E_{\mathbf{k_2}}}.
\eea
Finally, in the conditions $\lambda_1=-\lambda_4, \lambda_2=-\lambda_3$
we have:
\bea
F_{\lambda_1,\lambda_2,-\lambda_2,-\lambda_1}(\mathbf{k_1},\mathbf{k_2})=
\frac{1}{2} \frac{{k_1}{k_2}}{E_\mathbf{k_1}E_\mathbf{k_2}}\times \nonumber\\
\times\Bigg[1+ \frac{ \mathbf{k_1}\cdot\mathbf{k_2}\Big(\lambda_1\lambda_2 E_\mathbf{k_1}E_\mathbf{k_2}+m^2 \Big)}{k_1^2\, k_2^2}+\nonumber\\
+\frac{\im m  ( \mathbf{k_2}\times \mathbf{k_1}) \cdot\mathbf{u}_z (\lambda_1 E_\mathbf{k_1}+\lambda_2 E_\mathbf{k_2})}{k_1^2\, k_2^2} \Bigg]. \label{expab}
\eea
When $\lambda_1=-\lambda_4$ and $\lambda_2=\lambda_3$ we have
the following symmetry:
\bea
F^*_{\lambda_1\lambda_2\lambda_3\lambda_4} (\mathbf{k_1},\mathbf{k_2})&=&
{\phi^\dagger}_{\lambda_2}(\mathbf{k_2})
{\phi}_{\lambda_1}(\mathbf{k}_1)
{\phi^\dagger}_{\lambda_4}(\mathbf{k}_1),
{\phi}_{\lambda_3}(\mathbf{k_2})\nonumber\\
&=&F_{\lambda_2\lambda_1\lambda_4\lambda_3} (\mathbf{k_2},\mathbf{k_1})\,,
\eea
that is,  in expression (\ref{expab}) we have an identity upon the 
exchange of indexes $\lambda_1\leftrightarrow\lambda_2$, $\mathbf{k}_1\leftrightarrow\mathbf{k}_2$.

\section{Derivation of  Elliot's formula} \label{elliot_appendix}

The solution of the homogeneous problem presented in equation (\ref{bse_redux}) can be used to calculate the optical conductivity of the system.
We now detail the derivation of Elliot's formula for  TMDC's

First we decompose the excitonic wave function into a complete set of eingenfunctions of (\ref{bse_redux}):
\be
\Psi^{s\tau}_{\ell}(k)= \sum_{n} c^{s\tau}_{\ell n} \psi^{s\tau}_{\ell,n}(k)+ \int_0^\infty dq \,g^{s\tau}_\ell(q) \psi^{s\tau}_{\ell}(q,k), \label{eig_bse}
\ee
where we have separated the discrete and continuum states of the exciton spectrum, with $n$  and $q$ the radial quantum numbers, respectively.  We further recall the orthogonality relations: 
\bea
\int_0^\infty dk k[{\psi^{s\tau}_{\ell,n^\prime}}(k)]^\dagger\psi^{s\tau}_{\ell,n}(k)=\delta_{n^\prime,n}\\ \int_0^\infty dk k [{\psi^{s\tau}_{\ell,n}}(k)]^\dagger \psi^{s\tau}_\ell(q,k) \psi =0, \\\int_0^\infty dk k[{\psi^{s\tau}_\ell}(q^\prime,k)]^\dagger\psi^{s\tau}_\ell(q,k)=\delta(q-q^\prime),
\eea

The
non-homogeneous equation (\ref{bse_aprox}), after we substitute the expansion for $\Psi^{s\tau}_{\ell}(k)$ into the eigenfunctions  of 
the Kernel $K^{BS}$ (\ref{bse_matrix}), and integrating in $\theta$, becomes:
\bea
(\omega-K^{BS}_\ell)\left(\sum_{n} c_{\ell n}^{s\tau} \psi^{s\tau}_{\ell,n}(k)+ \int_0^\infty dq \,g^{s\tau}_\ell(q) \psi^{s\tau}_{\ell}(q,k) \right)\nonumber\\
=\frac{v^{s\tau}_{\ell,+}(k)}{\omega_{+}(k)}\,.
\eea
 where the angular decomposition of the velocity (\ref{velocity_matrix}),
\be
v^{s\tau}_{\ell,\lambda}(k)= \int_0^{2\pi} \frac{d\theta}{2\pi} v^{s\tau}_{\lambda}(\mathbf{k}) e^{\im(\ell+1)\theta},
\ee 
is composed of two terms:
\be
v^{s\tau}_{\ell,\lambda}= v^{s\tau}_{0,\ell} \delta_{0,\lambda}+ v^{s\tau}_{-2,\lambda} \delta_{-2,\ell},
\ee
remembering that the angular decomposition has an extra $\theta$ phase.
The  explicit expression for $v^{s\tau}_{\ell,\lambda}(k)$ reads:
\be
v^{s\tau}_{\ell,\lambda}(k)=\frac{1}{2} \left(\frac{m^{s\tau}}{E^{s\tau}_k}+(\ell+1) \frac{\lambda}{2} \right).
\ee

Using the eigenfunction orthogonality and neglecting the continuum part $\psi^{s\tau}_\ell(q,k)$, we arrive at:
\be
\sum_n(\omega-E^{s\tau}_{\ell n})c^{s\tau}_{\ell n}\psi^{s\tau}_{\ell,n}(k)= \frac{v^{s\tau}_{\ell,+}(k)}{\omega^{s\tau}_{+ }(k)}, \label{bse_aprox_2}
\ee
from where it follows the coefficients $c^{s\tau}_{\ell s}$:
\be
c_{\ell n}^{s\tau}= \int_0^\infty k dk \,{\psi^{s\tau}_{\ell,n}}^\dagger(k) \frac{v^{s\tau}_{\ell,+ }(k)}{\omega^{s\tau}_{+}(k) (\omega-E^{s\tau}_{\ell n})}\,.
\ee
Using the last result, the exciton contribution to the polarization is, using equation (\ref{pfinal}),  given by:
\bea
\frac{P}{S}&=&\sum_{s\tau, \ell=\{0,2\},n} \left|\int_0^\infty q dq \frac{v^{s\tau}_{\ell,+}(q)}{2E^{s\tau}_q   } \left[\psi^{s\tau}_{n\ell}(q) \right]^* \right|^2\times
\nonumber\\
 &\times&\frac{1}{ \omega-E^{s\tau}_{\ell n}+i\hbar\gamma} {\cal E}_0,
\eea
remembering here that we are using units such that  $v_F=\hbar=e=1$. 
Note that the $\theta$ integral has been performed, and we are summing over all the spin/valley indexes. If we use the definition of the weight function $M^{s\tau}_{n \ell}$ (\ref{weight_elliot}) we have:
\be
\frac{P}{S}= \sum_{s\tau, \ell=\{0,2\},n}  \frac{M^{s\tau}_{\ell n}}{\omega-E^{s\tau}_{\ell n}+i\gamma} {\cal E}_0,
\ee
where we have introduced a phenomenological relaxation rate $\gamma$.  The 2D optical susceptibility comes from $\mathbf{P}=\varepsilon_0\chi_{2D} {\cal \mathbf{E}}$:
\be
\chi_{2D}(\omega)= \frac{e^2}{\varepsilon_0}\sum_{s\tau, \ell=\{0,2\},n}  \frac{M^{s\tau}_{\ell n}}{\hbar\omega-E^{s\tau}_{\ell n}+i\hbar\gamma} ,
\ee
where we reintroduce the units, and the conductivity reads:
\be
\frac{\sigma(\omega)}{\sigma_0}= 4i \hbar\omega \sum_{s\tau, \ell=\{0,2\},n}  \frac{M^{s\tau}_{\ell n}}{\hbar\omega-E^{s\tau}_{\ell n}+i\hbar\gamma}\,. 
\label{elliot_cond}
\ee
Finally, the absorbance coefficient ${\cal A}(\omega)=1-{\cal T}(\omega)-{\cal R}(\omega)$, where ${\cal T}(\omega)$ [${\cal R}(\omega)$] is
the electromagnetic transmission [reflection] for a TEM wave, is given by:
\bea
{\cal A}(\omega)\approx \frac{\omega}{c\sqrt{\varepsilon_m}}\Im\left\{ \chi_{2D}(\omega) \right\}=\nonumber\\
\frac{4\pi\alpha\omega\gamma}{\sqrt{\varepsilon_m}}  \sum_{\sigma, \ell=\{0,2\},n}  \frac{M^{s\tau}_{\ell n}}{(\omega-E^{s\tau}_{\ell n}/\hbar)^2+\gamma^2} ,
\eea
where $\alpha\approx 1/137$ is the fine-structure constant.

\section{The Bethe-Salpeter kernel} \label{bethe_salpeter_kernel}

In this appendix we give the explicit forms of the BSE kernel. Firstly, from equation (\ref{excitonic_rabi}) we have:
\bea
{\cal B}^{s\tau}_{\mathbf{k}\lambda}(t)= \frac{1}{S}\sum_{\mathbf{q}} V(|\mathbf{k}-\mathbf{q}|)\Big[p^{s\tau}_{\lambda}(\mathbf{q},t) F^{s\tau}_{\lambda^\prime\lambda^\prime\lambda\lambda}(\mathbf{k},\mathbf{q})+\nonumber\\
+ p^{s\tau}_{\lambda^\prime}(\mathbf{q},t) F^{s\tau}_{\lambda^\prime \lambda \lambda^\prime\lambda}(\mathbf{k},\mathbf{q})\Big]\,.\label{excitonic_rabi_ap}
\eea
For the homogeneous case we only consider the first term 
in the previous equation and choose
with $\lambda=-\lambda'=+$, which corresponds to the resonant term. Thus we have the BSE kernel reading:
\be
K^{s\tau}_\lambda(k,q,\theta)= V(|\mathbf{k}-\mathbf{q}|)F^{s\tau}_{-\lambda-\lambda\lambda\lambda}(\mathbf{k},\mathbf{q})\,.
\ee
Using the expression (\ref{expab}) for the 
$F^{s\tau}_{--++}(\mathbf{k},\mathbf{q})$, and the Keldysh potential (\ref{Keldysh}), and after the angular decomposition 
\be
\tilde{T}^{s\tau}_\ell(k,q)= \int_0^{2\pi} \frac{d\theta}{2\pi} e^{\im(\ell+1)\theta}K^{s\tau}_+(k,q,\theta), \label{bse_ang_dec}
\ee
we have the corresponding kernel $\tilde{T}^{s\tau}_\ell(k,q)$:
\bea
\tilde{T}^{s\tau}_\ell(k,q)= -\frac{\alpha}{4 \pi\varepsilon_m} \frac{c}{v_F} \frac{kq^2}{2E_kE_q} \Big[ I_\ell(k,q)+\nonumber\\ c^{s\tau}_-(k,q)I_{\ell+1}(k,q)+c^{s\tau}_+(k,q) I_{\ell-1}(k,q) \Big], \label{t_aux}
\eea
where we have defined:
\be
I_\ell(k,q)= \int_0^{2\pi} d\theta \e^{\im(\ell+1)\theta} 
\frac{q_0}{p_{k,q}(\theta)(p_{k,q}(\theta)+q_0)},  \label{int_func}
\ee
\bea
c^{s\tau}_\pm(k,q)&=&\frac{1}{2kq}\left[E^{s\tau}_k E^{s\tau}_q+ m_{s\tau}^2
\right.
\nonumber\\
&\pm&\left. m_{s\tau}(E^{s\tau}_k+E^{s\tau}_q) \right],
\eea
and
\be
p_{k,q}(\theta)=\sqrt{q^2+k^2-2kq\cos\theta},
\ee
where $\alpha\approx1/137$ is the fine structure constant and
 $c$ is the speed of  light. 

For the kernel in the non-homogeneous BSE, we write the renormalization of  Rabi frequency  as:
\bea
{\cal B}^{s\tau}_{\mathbf{k}\lambda}(t)=  \int \frac{d\mathbf{q}}{(2\pi)^2} V(|\mathbf{k}-\mathbf{q}|)F^{s\tau}_{-\lambda-\lambda\lambda\lambda}(\mathbf{k},\mathbf{q}) p^{s\tau}_{\lambda}(\mathbf{q},\omega) +\nonumber\\
\int \frac{d\mathbf{q}}{(2\pi)^2} V(|\mathbf{k}-\mathbf{q}|)F^{s\tau}_{-\lambda \lambda-\lambda\lambda}(\mathbf{k},\mathbf{q}) p^{s\tau}_{-\lambda}(\mathbf{q},\omega)\,. 
\eea
Thus, we have  two kernels to consider:
\be
K^{1,s\tau}_\lambda=V(|\mathbf{k}-\mathbf{q}|)F^{s\tau}_{-\lambda-\lambda\lambda\lambda}(\mathbf{k},\mathbf{q}),
\ee
\be
K^{2,s\tau}_\lambda=V(|\mathbf{k}-\mathbf{q}|)F^{s\tau}_{-\lambda \lambda-\lambda\lambda}(\mathbf{k},\mathbf{q})\,.
\ee
Note that, in this case and contrary to the homogeneous BSE, we have to keep both terms in the renormalization of 
the Rabi frequency, as otherwise the real part of the optical conductivity would not have the correct positive sign. That is because both the resonant and off-resonance terms contribute to the optical conductivity, as is well known
in the non-interacting case.
Proceeding  as before,  the  angular decomposition (\ref{bse_ang_dec})
of the kernels leads to:
\bea
T^{1/2,s\tau}_{\lambda,\ell}(k,q)=  {\cal C}^{s\tau}_{k,q}\,  \Big[ I_\ell(k,q)+c^{1/2,s\tau}_{\lambda-}(k,q)I_{\ell+1}(k,q)+\nonumber\\
+c^{1/2,s\tau}_{\lambda+}(k,q) I_{\ell-1}(k,q) \Big],
\eea
with
\be
{\cal C}^{s\tau}_{k,q}=-\frac{\alpha}{4 \pi\varepsilon_m}\frac{c}{v_F}\frac{kq^2}{2E^{s\tau}_kE^{s\tau}_q},
\ee
and
\bea
c^{1,s\tau}_{\lambda\pm}(k,q)&=&\frac{1}{2kq}\left[E^{s\tau}_k E^{s\tau}_q+m_{s\tau}^2\pm \lambda m_{s\tau}(E^{s\tau}_k
\right.
\nonumber\\
&+&\left.E^{s\tau}_q) \right],
\eea
\bea
c^{2,s\tau}_{\lambda\pm}(k,q)&=&\frac{1}{2kq}\left[-E^{s\tau}_k E^{s\tau}_q+m_{s\tau}^2\pm \lambda m_{s\tau}(E^{s\tau}_k
\right.\nonumber\\
&-&\left.E^{s\tau}_q) \right],
\eea
where $\lambda=\pm$ and $\lambda^\prime=-\lambda$, and $I_{\ell}(k,q)$ is defined in equation (\ref{int_func}). 

The matrix element of the velocity operator appearing as the source term in the non-homogeneous BSE, was calculated in the previous section and reads:
\be
v^{s\tau}_{\ell,\lambda}(k)=\frac{1}{2} \left(\frac{m_{s\tau}}{E^{s\tau}_k}+(\ell+1) \frac{\lambda}{2} \right).
\ee
This concludes the presentation of the mathematical steps leading to the BSE equation discussed in the main text.

\section{Exchange correction around the $\Gamma$ point}

From equation (\ref{useful}), the exchange self-energy correction to
the transition energy between bands $i$ and $j$ is:
\bea
\Sigma^{xc}_{ij}(\mathbf{k})&=S\sum_{\mathbf{q}} V(\mathbf{q}) \sum_\lambda n_\lambda(\mathbf{k}-\mathbf{q})\times \nonumber\\
&\left[ F_{i\lambda\lambda i}(\mathbf{k},\mathbf{k}-\mathbf{q})-F_{j\lambda\lambda j}(\mathbf{k},\mathbf{k}-\mathbf{q})\right].
\eea
Neglecting temperature and doping effects, for the three band-model of reference \cite{Liu2013threeband}, the
only term that contributes to the exchange self-energy is the one with $\lambda=0$ (the valence band):
\bea 
\Sigma^{xc}_{ij}(\mathbf{q})=S\sum_{\mathbf{q}} V(\mathbf{q})
\left[F_{i00i}(\mathbf{k},\mathbf{k}-\mathbf{q})\right.-
\nonumber\\
\left.
F_{j00j}(\mathbf{k},\mathbf{k}-\mathbf{q})\right]. \label{exgamma}
\eea
and we make $i=1,2$ and $j=0$.  

\begin{figure}
	\centering
	\includegraphics[scale=0.5]{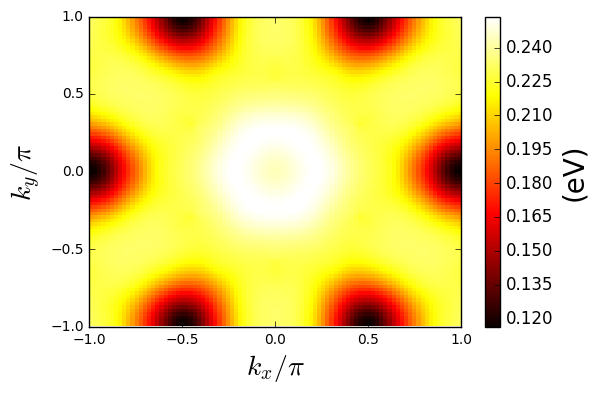}
	\\
	\includegraphics[scale=0.5]{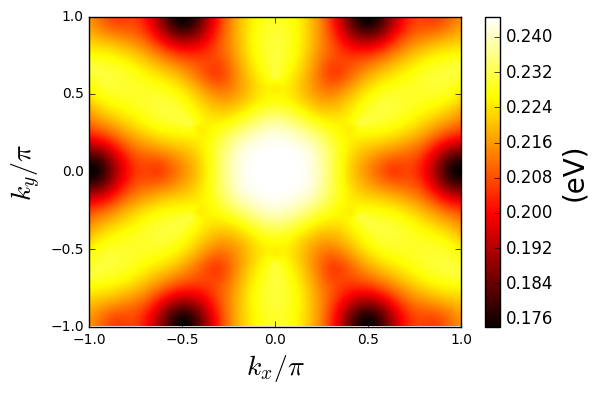}
	\caption{Exchange energy over the Brillouin zone for the $j=0$ to $i=1$ transition (top) and for the $j=0$ to $i=2$ transition (bottom).
		Note the six-fold symmetry of the exchange, which was computed using the three-band tight-binding model.
	} \label{exchange}
\end{figure}

To remove  the integrable divergence at $q=0$, that comes from the Keldysh potential, we
use polar coordinates leading to the need of computing the following integral
\bea
\Sigma^{xc}_{i0}(\mathbf{k})=-\sum_{j=1}^6
\int_{j\pi/3}^{(j+1)\pi/3} \frac{d\theta}{2\pi} \int_0^{q_0\sec(\theta-\pi/6-j\pi/3)}  \frac{qdq}{2\pi} 
\nonumber\\
\times
V(q)Q_i(\mathbf{k},q,\theta)
\eea
where $Q_i(\mathbf{k},q,\theta)= F_{i00i}(\mathbf{k},\mathbf{k}-\mathbf{q})-F_{0000}(\mathbf{k},\mathbf{k}-\mathbf{q})$ and $q_0$
reads
\begin{equation}
	q_0=\frac{2\pi}{a_0\sqrt 3}\,.
\end{equation}
In figure \ref{exchange} we show the exchange energy in the Brillouin zone for the $j=0$ to $i=1,2$ transitions, for the three-band tight-binding model and for the parameters of MoS$_2$.
The magnitude of the correction to the bare bands is of the order of 2.1-2.6 eV, which is
a substantial value.

\end{appendix}


\section*{Bibliography}
\bibliographystyle{iopart-num}
 \providecommand{\newblock}{}

\end{document}